\documentclass[3p,preprint,authoryear]{elsarticle}

\usepackage{graphicx}
\usepackage{amssymb}
\usepackage{amsmath}
\usepackage{bbm} 
\usepackage{lineno}

\modulolinenumbers[5]
\usepackage{hyperref}
\hypersetup{colorlinks=true,citecolor=blue}
\usepackage{color}
\usepackage{multirow}
\usepackage{booktabs}
\usepackage{bm}
\usepackage{epstopdf}
\usepackage{upgreek}
\usepackage{subfigure}
\usepackage{appendix}
\usepackage{setspace}
\usepackage{amsmath}

\usepackage{framed} 
\usepackage{multicol} 
\usepackage{nomencl} 
\renewcommand*{\nompreamble}{8pt}
\makenomenclature
\renewcommand*\nompreamble{\begin{multicols}{2}}
\renewcommand*\nompostamble{\end{multicols}}
\usepackage[font=small,labelfont=bf,labelsep=space]{caption}
\begin{document}

\begin{frontmatter}


\title{Discrete dislocation plasticity HELPs understand hydrogen effects in bcc materials}


\author[materials]{H. Yu}
\ead{haiyang.yu@materials.ox.ac.uk}  

\author[engineering]{A.C.F. Cocks}
\author[materials]{E. Tarleton \corref{cor}}
\ead{edmund.tarleton@materials.ox.ac.uk} 

\cortext[cor]{Corresponding author}
\address[materials]{Department of Materials, University of Oxford, Parks Road, OX1 3PH, UK}
\address[engineering]{Department of Engineering Science, University of Oxford, Parks Road, OX1 3PJ, UK}

\begin{abstract}
In an attempt to bridge the gap between atomistic and continuum plasticity simulations of hydrogen in iron, we present three dimensional discrete dislocation plasticity simulations incorporating the hydrogen elastic stress and a hydrogen dependent dislocation mobility law. The hydrogen induced stress is incorporated following the formulation derived by \cite{Gu2018} which here we extend to a finite boundary value problem, a microcantilever beam, via the superposition principle. The hydrogen dependent mobility law is based on first principle calculations by \cite{Katzarov2017} and was found to promote dislocation generation and enhance slip planarity at a bulk hydrogen concentration of 0.1 appm; which is typical for bcc materials. The hydrogen elastic stress produced the same behaviour, but only when the bulk concentration was extremely high. In a microcantilever, hydrogen was found to promote dislocation activity which lowered the flow stress and generated more pronounced slip steps on the free surfaces. These observations are  consistent with the hydrogen enhanced localized plasticity (HELP) mechanism, and it is concluded that both the hydrogen elastic stress and hydrogen increased dislocation mobility are viable explanations for HELP. However it is the latter that dominates at the low concentrations typically found in bcc metals.
\end{abstract}


\end{frontmatter}

\section{Introduction}
\label{intro}

Hydrogen can cause premature failure accompanied by a decrease in ductility in metals, which is often referred to as hydrogen embrittlement (HE) \citep{Ayas2014}. The underlying mechanism of this phenomenon has been under debate for the past two decades and several theories have been proposed \citep{Robertson2015}. For non hydride forming material, the prevailing mechanisms are hydrogen enhanced decohesion (HEDE) and hydrogen enhanced localized plasticity (HELP). The HEDE mechanism assumes that dissolved hydrogen reduces the cohesive strength of the material lattice \citep{Oriani1972}. It is supported by atomistic simulations where cohesive strength decreases with increasing hydrogen content \citep{Lynch2012}. The HELP mechanism assumes that hydrogen facilitates dislocation activity thereby enhancing plasticity locally \citep{Birnbaum1994}, which is backed up by experimental observations and theoretical calculations \cite{Robertson2001}. In recent years, it has become widely accepted that HE is a complex,  material and environmental dependent process \citep{Robertson2015} so that no mechanism applies exclusively. In fact, it has been indicated that the aforementioned mechanisms may act in concert to promote failure \citep{Novak2010}. \citet{Barrera2014} performed continuum level HE simulation considering both mechanisms in welded structures, where microcracks formed at the matrix/carbide interface due to decohesion process followed by localization of plastic flow responsible for the link-up of the microcracks. Most recently, a novel hydrogen-enhanced-plasticity mediated decohesion mechanism was proposed for HE in martensitic steels \citep{Nagao2018,Nagao2012}. It claims that hydrogen-enhanced mobility of dislocations leads to local stress build-up and hydrogen concentration elevation close to material interfaces where decohesion is promoted. In this mechanism, hydrogen enhanced plasticity is the driving force for HE, which highlights the importance of understanding how hydrogen affects dislocations.

\subsection{Experimental evidence}
Experimental evidence for the HELP mechanism is found in the literature over different length scales. In situ deformation experiments using an environmental cell inside a transmission electron microscope (TEM) \citep{Ferreira1998} enables direct observation of hydrogen-dislocation interactions involving several dislocation lines. With this technique, \citet{Shin1998} observed hydrogen enhanced dislocation motion in $\alpha$-Titanium, which was eliminated and recovered when the hydrogen environment was removed and reintroduced. A semi-quantitative relationship between dislocation velocity and hydrogen gas pressure was also extracted. Similar experiments were performed on a variety of materials, including Fe \citep{Tabata1983}, Ni \citep{Robertson1986}, Al \citep{Bond1988} and their alloys \citep{Rozenak1990,Bond1989}. A surprising feature of these studies is that the enhancement in the mobility of dislocations due to hydrogen is independent of crystal structure and is the same for edge, screw, mixed, and partial dislocations \citep{Robertson2001}. Using a similar technique, \citet{Ferreira1998} observed hydrogen decreased the equilibrium separation in a dislocation pileup in $310$S stainless steel and Al; indicating hydrogen has a shielding effect on dislocation interactions. \citet{Ferreira1999} also observed hydrogen suppressed dislocation cross-slip in Al and enhanced slip planarity. 

Plasticity is an emergent property due to the collective behavior of a large number of dislocations. Therefore it cannot be fully understood by only studying a small number. Small scale tests such as nanoindentation \citep{Nibur2006,Afrooz2010} and microcantilever bending \citep{Tarlan2017,Yun2017,Yun2018} in hydrogen environments provide essential new insight. In these experiments, the global load-displacement curve is recorded, as well as dislocation pileups and plastic localisation by microscopy techniques. \cite{Nibur2006} studied hydrogen effects on dislocation activity in austenitic stainless steel with nanoindentation. Hydrogen was observed to facilitate dislocation nucleation and enhance slip planarity. \cite{Tarlan2017} performed in situ bend tests on notched microcantilevers of single crystal Fe-$3$ wt$\%$ Si alloy and observed that hydrogen reduced flow stress and promoted crack propagation. \cite{Yun2017} performed similar experiment on un-notched microcantilevers of single crystalline FeAl and observed that hydrogen reduced the flow stress and enhanced slip localisation. Recently, \cite{Yun2018} investigated hydrogen assisted fracture in single crystalline FeAl notched microcantilevers, which demonstrated that hydrogen leads to global softening and reduced fracture toughness. It should be noted that the decrease in flow stress in these experiments was explained with hydrogen enhanced dislocation generation and the accelerated crack propagation with hydrogen reduced dislocation mobility. Ex situ methods where specimens are loaded in a hydrogen environment and observed at specific locations are also widely adopted, and allow tests to be performed at a larger scale. Four point bend tests on lath martensitic steel were carried out \citep{Nagao2012}, after which the fracture surface was examined with scanning electron microscopy (SEM) and the microstructure immediately beneath the prominent features on the fracture surface was determined by focused-ion beam (FIB) machining and transmission electron microscopy (TEM). Significant plasticity, consistent with the HELP mechanism, was observed beneath hydrogen induced \lq\lq quasi-cleavage \rq\rq fracture surface, which inspired the hydrogen-enhanced-plasticity mediated decohesion mechanism \citep{Nagao2018}. A wealth of experimental evidence on the macroscopic scale is also found in the literature, for instance, intensive localized shear was observed on fracture surfaces in hydrogen charged AISI $310$S stainless steel \citep{Ulmer1991}, and hydrogen was observed to change microvoid failure mode from inter-ligament necking to shear banding in slow strain rate tensile test \citep{Takashi2014}.

\subsection{Previous modeling and simulation}
\label{intro:model}
The experimental evidence for the HELP mechanism demonstrates two different hydrogen effects;  decreasing the equilibrium dislocation spacing while simultaneously increasing the dislocation mobility. To some extent, this can be accounted for by the elastic shielding theory \citep{Birnbaum1994}. In this theory, hydrogen forms an atmosphere around the dislocations. This generates a first order elastic interaction arising from introducing hydrogen atoms within the dislocation stress field and a second-order interaction resulting from a local change in the elastic modulus caused by solute hydrogen. It was shown by \citet{Birnbaum1994} that these elastic interactions led to a short-range decrease in the elastic force experienced by a dislocation due to another dislocation or an obstacle. In this way, hydrogen decreases dislocation spacing by shielding dislocation interactions and enhances dislocation mobility by shielding elastic stress fields of other obstacles. At very high background concentrations such as $c_0=1\times10^4$ appm (equivalent with 1 hydrogen atoms per 100 solvent atoms) the hydrogen elastic shielding effect can be considerable. Within the same theoretical framework, the experimentally observed enhanced slip planarity was explained by \citet{Ferreira1999}. In fcc Al, the first order elastic interaction dominates, leading to the formation of hydrogen atmospheres primarily around edge components. In this way, the self-energy of an edge dislocation-hydrogen complex is lowered while that of the screw component is less affected. This makes the edge components more favorable when compared to the hydrogen free case, therefore, the proportion of screw component will be decreased and the tendency to cross-slip suppressed with hydrogen. This mechanism is shown to be noticeable with a hydrogen concentration of $c_0>1\times10^4$ appm. Strictly speaking, the elastic shielding theory is based on an implicit assumption that hydrogen diffuses fast enough that the hydrogen distribution is always in equilibrium, in terms of potential, with a remote reservoir with a uniform background concentration $c_0$. In reality, $c_0$ should be taken as the bulk concentration \citep{Jagodzinski2000,Wen2003}. It is established that hydrogen diffusivity in bcc metals can be orders of magnitude higher than in fcc metals at ambient temperatures \citep{Lynch2012}, therefore, this theory should be more applicable to bcc metals \citep{Gu2018}. However, the low hydrogen solubility in bcc metals gives a small bulk concentration $c_0<1.0$ appm, at which hydrogen elastic shielding is negligible. Even for fcc metals, the concentration used in \citet{Birnbaum1994} is too high. This implies that the hydrogen elastic shielding mechanism, while being viable, might not be the dominant factor in HELP phenomena, as noted by other researchers \citep{Teus2007,Taketomi2008,Gavriljuk2010} and will be revisited in this paper.

Atomistic and quantum mechanics based simulations are increasingly utilized to simulate hydrogen-dislocation interactions, providing fundamental insights and sometimes controversial evidence for the HELP mechanism. Using molecular dynamics, \cite{Song2011,Song2013} observed hydrogen suppressed dislocation emission from a crack tip which facilitated brittle-cleavage failure in $\alpha$-Fe. They also studied the effect of hydrogen on a pure edge dislocation in $\alpha$-Fe and concluded that hydrogen reduces dislocation mobility over a wide range of concentrations \citep{Song2014}, consistent with the solute drag theory and in contraction with the HELP mechanism. A similar scenario was observed by \cite{Bhatia2014} where hydrogen  impeded the motion of pure edge dislocations in $\alpha$-Fe. According to \cite{Song2014}, these results are also applicable to fcc metals due to certain similarities of edge dislocations between the two systems. However, the story is quite different when it comes to screw dislocations. In bcc metals, the mobility of screw dislocations is much lower than for edge. Plasticity is therefore mainly limited by screw dislocations in bcc metals whose mobility is dominated by kink-pair nucleation and migration \citep{anderson2017}. Using a line tension model, \cite{Itakura2013} studied hydrogen effect on screw dislocation mobility in $\alpha$-Fe using first-principles calculation and revealed a general trend that hydrogen enhances screw mobility by promoting nucleation of kink pairs at a low concentration and decreases the mobility by impeding migration of kink pairs at a high concentration. Based on this work and by performing kinetic Monte Carlo (kMC) simulations, \cite{Katzarov2017} were able to simulate the motion of individual $1/2[111]$ screw dislocations over long timescales. They provided the first quantitative description of hydrogen effects on screw mobility as a function of bulk hydrogen concentration, temperature and applied stress in $\alpha$-Fe. According to this study, hydrogen significantly enhances screw mobility up to a bulk concentration $c_0=5$ appm at room temperature, which indicates the HELP mechanism takes place in $\alpha$-Fe given the small solubility of hydrogen. In their multiscale quantum-mechanics/molecular mechanics simulation, \cite{Zhao2011} observed hydrogen lowered the Peierls energy barrier for screw dislocations significantly, indicating enhanced mobility. Comparing this conclusion to their observation of hydrogen decreased mobility in edge dislocations, \cite{Bhatia2014} proposed that hydrogen has a tendency to make the crystal more isotropic in terms of dislocation motion, i.e. increasing screw mobility while decreasing edge mobility \citep{Itakura2013}, which was also hypothesized by \cite{Moriya1979} in an experimental investigation.

Continuum level simulations to investigate HELP  have been performed and provided insight into hydrogen induced failure at the macroscopic scale. \cite{Sofronis2001} and \cite{Liang2003} showed that hydrogen could induce shear instability in a plain strain tensile specimen. \cite{Ahn2007} observed hydrogen enhanced void growth and coalescence using a representative material volume approach. \cite{Barrera2016} studied hydrogen effects on plastic behavior of notched tensile plates using a coupled mechanical-diffusion analysis. A similarity of these studies was that the HELP mechanism was incorporated by describing the matrix yield stress as a linearly decreasing function of local hydrogen concentration. Most recently, \cite{Yu2018} adopted a sigmoidal relationship for the hydrogen softening effect to study hydrogen-microvoid interactions and concluded that the results were more comparable to certain experimental observations. Nevertheless, both the linear relation and the sigmoidal relation are attempts to describe the HELP mechanism phenomenologically, due to the lack of quantification of the hydrogen softening effect applicable to the continuum scale, either experimentally or numerically. Although many clues about the physics of hydrogen-dislocation interactions have been revealed by atomistic simulations, they cannot be directly utilized in continuum level studies due to the large gap in spatial and temporal scales. 

To bridge this gap, discrete dislocation dynamics (DDD) simulations have proven to be an effective tool. DDD can simulate the aggregate behavior of large dislocation ensembles by simulating the dynamics and interactions of every discrete dislocation segment. DDD has been used to study a wide range of problems assuming isotropic elasticity such as the formation and strength of dislocation junctions \citep{Capolungo2011,Wu2013} and work hardening \citep{Arsenlis2007}. A small number of studies have also been performed using anisotropic elasticity to simulate a Frank-Read source \citep{Fitzgerald2012} and dislocation loops \citep{Aubry2011}. DDD can be coupled with  FEM to simulate a finite domain with mixed boundary conditions; this is usually referred to as discrete dislocation plasticity or DDP. This approach can help understand plasticity due to the collective behavior of a large number of dislocations in a confined volume under various loading scenarios, such as in a micro-beam under bending \citep{Motz2008} or a micropillar under tension \citep{Motz2009} or compression \citep{Ryu2013}. In these simulations, the dynamics of individual dislocation segments and their interactions are described in a manner consistent with the physics at the atomistic scale, and the results reflect crystal level plasticity. Within the DDP framework, environmental effects on plasticity can be studied by properly considering the effect of environment on dislocation dynamics or dislocation-obstacle interactions. By incorporating a temperature dependent dislocation mobility law informed by atomistic simulation \citep{Po2016}, \cite{Cui2016} studied the effect of temperature on plasticity in bcc micropillar crystals. \cite{Cui2017} also studied the effect of irradiation on strain bursts in fcc and bcc single crystals by considering the irradiation effects on initial dislocation structures. 

Hydrogen was recently incorporated in three dimensional DDD by \cite{Gu2018}, which is a promising method for bridging the gap in the multi-scale simulation of the HELP mechanism. They simulated the role of hydrogen during the shrinking of a dislocation loop and in the arrangement of parallel dislocations. Hydrogen atoms were regarded as point defects, misfitting inclusions in an elastic volume, subjected to the externally applied stress and the dislocation stress fields. The elastic stress field due to hydrogen was derived in three dimensional space, following the approach proposed by \cite{Cai2014}. In this way, the hydrogen elastic energy contribution to dislocation activity was considered. In the case with a high hydrogen diffusivity, e.g. in bcc materials, it was shown that hydrogen tended to compress the dislocation loop in the directions parallel to the Burgers vector and to decrease the spacing of parallel edge dislocations, consistent with hydrogen elastic shielding theory. The work of \cite{Gu2018} is therefore the first three dimensional DDD illustration of the hydrogen elastic shielding mechanism in an infinite medium. DDP simulations of the effect of hydrogen on dislocation plasticity in a confined volume has not been reported in the literature previously. Further, it is noted that the influence of hydrogen in the previous DDD simulations was noticeable only at very high bulk concentrations $c_0\approx5\times10^4$ appm and $c_0\approx1\times10^5$ appm. This is a consequence of only considering hydrogen elastic shielding and therefore indicates that other hydrogen effects need to be taken into account in order to reproduce the influence of hydrogen on plasticity at the low concentrations observed in bcc materials.
%
%

\subsection{The present work}
In this paper, we present three dimensional discrete dislocation plasticity (DDP) simulations incorporating hydrogen in a finite volume with dislocation mobility parameters typical for bcc metals and a physical bulk hydrogen concentration. In addition to hydrogen elastic shielding, hydrogen increased dislocation mobility is also incorporated following the quantitative description introduced by \citet{Katzarov2017}. The simulations were performed using an in-house Matlab based GPU accelerated DDP code; which is a modified version of \emph{DDLab} developed at LLNL by \citet{Arsenlis2007}. 

The micro-cantilever geometry, popular for experimental studies on hydrogen embrittlement, is simulated using the superposition method \citep{Giessen1999,Jaafar2008}. In contrast to \citet{Gu2018} where dislocation mobilities for edge and screw components were assumed equal, here we assign a higher mobility to edge components than  screw, which is more representative of bcc metals. With a hydrogen dependent mobility law, a reduction in the global reaction force (which is proportional to the flow stress) was observed even with a low bulk hydrogen concentration, typical of bcc metals. Further, it is concluded that the contribution from hydrogen elastic shielding is negligible unless the concentration is artificially high. In addition, it is shown that hydrogen can possibly increase slip planarity by decreasing the proportion of pure screw segments, consistent with experimental observations. 

\section{Methodology}
\label{method}
\subsection{Peach-Koehler force}
\begin{figure}[!h]
\centering\includegraphics[width=0.5\linewidth]{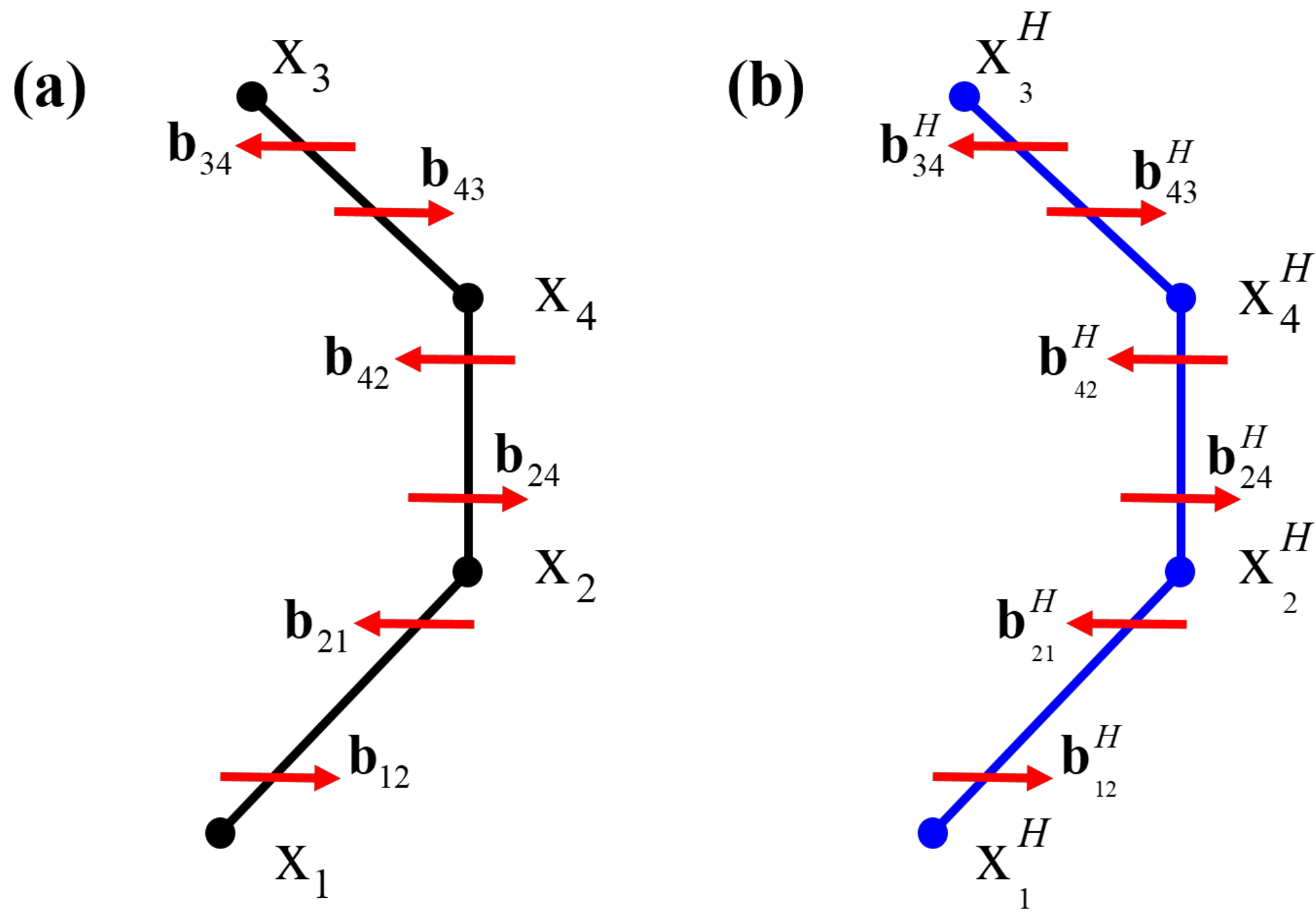}
\caption{Illustration of (a) dislocation segments in DDD simulation and (b) hydrogen-related dislocation segments, adapted from \citet{Gu2018}.}
\label{fig:segs}
\end{figure}

The DDD simulation in this work are nodal based \citep{Cai2006book,Arsenlis2007}, i.e. dislocations are represented by discrete straight segments as shown in \autoref{fig:segs}(a), and the dislocation motion is accounted for by evaluating the velocity $\bm{V}_k$ of node $k$ at position $\bm{X}_k$ through a general linear mobility law which describes the dislocation mobility with various modes such as glide, climb and cross-slip.

In general, the force per unit length on a dislocation line segment bounded by nodes $k$ and $l$ is
\begin{equation}
\label{eq:pk1}
\bm{f}_{kl}(\bm{x}) = \left(\bm{\sigma}(\bm{x})\cdot \bm{b}_{kl} \right)\times\bm{l}_{kl},
\end{equation}
where $\bm{\sigma}(\bm{x})$ is the local stress at point $\bm{x}$ on the dislocation line, $\bm{b}_{kl}$ is the Burgers vector, and $\bm{l}_{kl}$ is the unit tangent vector of the dislocation line segment. In the DDD simulation, the non-singular continuum theory of dislocations \citep{Cai2006} is adopted. Using the convention that nodal values are denoted in uppercase, the total nodal force on node $k$, connected to node $l$ is evaluated in three parts
\begin{equation}
\label{eq:Fk}
\bm{F}_{k} = \sum_{l}\sum_{i,j}\tilde{\bm{f}}^{ij}_{kl}(\bm{X}_k)  + \hat{\bm{f}}(\bm{X}_k) + \bm{f}^c(\bm{X}_k),
\end{equation}
where $\tilde{\bm{f}}^{ij}_{kl}$ is the elastic force due to segment $i\rightarrow j$ integrated along segment $k\rightarrow l$. This is summed over all segments $i\rightarrow j$ inside the domain, including the self force due to segment $k\rightarrow l$, this is then summed over all nodes $l$ which are connected to node $k$. $\hat{\bm{f}}$ is the corrective elastic force obtained using FEM to account for the applied and image stresses and $\bm{f}^c$ is the force due to the dislocation core energy.

The PK force due to dislocations arises from the stress fields of straight dislocation segments. Take the configuration in \autoref{fig:segs}(a) for instance, the stress tensor at a field point $\bm{x}$ induced by the dislocation segment $1\rightarrow 2$ can be expressed as \citet{Arsenlis2007}
\begin{equation}
\label{eq:disstress}
\begin{aligned}
\tilde{\bm{\sigma}}^{12}(\bm{x}) = &-\frac{\mu}{8\pi}\int_{\bm{X}_1}^{\bm{X}_2}\left(\frac{2}{R_a^3}+\frac{3a^2}{R_a^5}\right)\left[(\bm{R}\times\bm{b}_{12})\otimes \mathrm{d}\bm{x'}+\mathrm{d}\bm{x'}\otimes(\bm{R}\times\bm{b}_{12})\right]\\
&+\frac{\mu}{4\pi(1-\nu)}\int_{\bm{X}_1}^{\bm{X}_2}\left(\frac{1}{R_a^3}+\frac{3a^2}{R_a^5}\right)\left[(\bm{R}\times\bm{b}_{12})\cdot \mathrm{d}\bm{x'}\right]\bm{I}\\ 
&-\frac{\mu}{4\pi(1-\nu)}\int_{\bm{X}_1}^{\bm{X}_2}\frac{1}{R_a^3}\left[(\bm{b}_{12}\times \mathrm{d}\bm{x'})\otimes\bm{R}+\bm{R}\otimes(\bm{b}_{12}\times \mathrm{d}\bm{x'})\right]\\
&+\frac{\mu}{4\pi(1-\nu)}\int_{\bm{x}_1}^{\bm{x}_2}\frac{3}{R_a^5}\left[(\bm{R}\times\bm{b}_{12})\cdot \mathrm{d}\bm{x'}\bm{R}\otimes\bm{R}\right],
\end{aligned}
\end{equation}
where $\mu$ is the shear modulus, $\nu$ the Poisson's ratio, $a$ the core width and $\bm{I}$ the second order identity tensor. $\bm{R}=\bm{x}-\bm{x'}$, and $R_a=\sqrt{\bm{R}\cdot\bm{R}+a^2}$. The nodal force at node $3$ due to the interaction between segment $4\rightarrow3$ with Burgers vector $\bm{b}_{43}$, and the segment $1\rightarrow 2$ with stress field $\tilde{\boldsymbol{\sigma}}^{12}$ is
\begin{align}
\label{eq:pkdisseg}
\tilde{\bm{F}}_{3}=\tilde{\bm{f}}^{12}_{43}(\bm{X}_3) & = \int_0^{L_{43}}\frac{l}{L_{43}}\left(\tilde{\bm{\sigma}}^{12}\left[\bm{x}(l)\right]\cdot\bm{b}_{43}\right)\times\bm{l}_{43}\mathrm{d}l\\
\textrm{where}\quad\bm{x}(l)&=\left(1-\frac{l}{L_{43}}\right)\bm{X}_4+\frac{l}{L_{43}}\bm{X}_3, \nonumber
\end{align}
$L_{43}=\lvert \bm{X}_3-\bm{X}_4 \rvert$, and $\bm{l}_{43}=(\bm{X}_3-\bm{X}_4)/L_{43}$. Analytic expressions for \autoref{eq:pkdisseg} have been given by \citet{Arsenlis2007}.

\subsection{Dislocation mobility law}

A linear dislocation mobility law is adopted. For each segment $ij$, a drag tensor $\bm{B}_{ij}$ is determined according to the segment character, the nodal velocity $\bm{V}_k$ at node $k$ is then obtained using
\begin{equation}
\label{eq:linearmob}
\left[\frac{1}{2}\sum_{l}L_{kl}\bm{B}_{kl}\right]^{-1}\bm{F}_{k} = \bm{V}_k,
\end{equation}
where the sum is again over all nodes $l$ connected to node $k$, and $\bm{F}_k$ is the nodal force determined by \autoref{eq:Fk}. To construct drag tensor suitable for bcc materials \citep{Cai2006book}, the cases of pure screw and pure edge components are considered first. A pure screw dislocation segment is assigned with isotropic mobility in all directions perpendicular to its line direction, which mimics the so-called \lq\lq pencil-glide\rq\rq
\begin{equation}
\label{eq:dragscrew}
\bm{B}_{kl}(\bm{l}_{kl}) = B_s(\bm{I}-\bm{l}_{kl}\otimes\bm{l}_{kl})+ B_{l}(\bm{l}_{kl}\otimes\bm{l}_{kl}),\quad \mathrm{when}\ \bm{l}_{kl}\parallel \bm{b}_{kl},
\end{equation}
where $B_s$ is a drag coefficient for the motion of screw dislocations and $B_l\ll B_s$ to allow a node to move rapidly along the line direction. Meanwhile, the mobility of a pure edge segment is anisotropic with respect to glide and climb,
\begin{equation}
\label{eq:dragedge}
\bm{B}_{kl}(\bm{l}_{kl}) = B_{eg}(\bm{m}_{kl}\otimes\bm{m}_{kl})+B_{ec}(\bm{n}_{kl}\otimes\bm{n}_{kl})+B_{l}(\bm{l}_{kl}\otimes\bm{l}_{kl}),\quad \mathrm{when}\ \bm{l}_{kl}\perp\bm{b}_{kl},
\end{equation}
where $B_{eg}$ and $B_{ec}$ are the drag coefficients for glide and climb, respectively. The unit vectors are the plane normal $\bm{n}_{kl} = (\bm{b}_{kl}\times\bm{l}_{kl})/\lvert \bm{b}_{kl}\times\bm{l}_{kl}\rvert$ and glide direction $\bm{m}_{kl} = \bm{n}_{kl}\times\bm{l}_{kl}$. To account for the mobility of mixed segments, an interpolation function was proposed by \citet{Cai2004}, enabling a smooth transition between the two limits
\begin{equation}
\begin{aligned}
\label{eq:Bkl}
\bm{B}_{kl}(\bm{l}_{kl}) &= \lvert \bm{b}_{kl}\rvert\left[B^{-2}_{eg}\lvert\bm{b}_{kl}\times\bm{l}_{kl}\rvert^2+B^{-2}_{s}(\bm{b}_{kl}\cdot\bm{l}_{kl})^2\right]^{-1/2}(\bm{m}_{kl}\otimes\bm{m}_{kl})\\
&+ \frac{1}{\lvert \bm{b}_{kl}\rvert}\left[B^{2}_{ec}\lvert\bm{b}_{kl}\times\bm{l}_{kl}\rvert^2+B^{2}_{s}(\bm{b}_{kl}\cdot\bm{l}_{kl})^2\right]^{1/2}(\bm{n}_{kl}\otimes\bm{n}_{kl})
+B_{l}(\bm{l}_{kl}\otimes\bm{l}_{kl}).
\end{aligned}
\end{equation}
The dislocation mobility is inversely proportional to the drag coefficient. In bcc materials, the mobility of a pure edge segment should be greater than a pure screw segment, which is accounted for by assigning $B_{eg} < B_s$. Climb is neglected throughout this work by using a $B_{ec}\gg B_{s}$. More details regarding the selection of drag coefficients are presented in \autoref{DDDmodel}. Besides, it should be noted that the \lq\lq pencil glide\rq\rq assumption for pure screw mobility is a simplified treatment, however, it is sufficient for the present work where the effect of hydrogen on slip planarity is attributed to its influence on the proportion of near screw segments and this effect will remain qualitatively unchanged with a different cross-slip law.

\subsection{Treatment of finite boundary conditions}
\label{BCwithoutH}

Dislocations in a finite medium are subjected to the corrective force term $\hat{\bm{f}}$ which appears on the right hand side of \autoref{eq:Fk}. To evaluate the corrective fields the superposition principle \citep{Needleman1995,Weygand2002} is adopted. Denoting the infinite-body fields as $\tilde{()}$ and the finite-element correction fields with $\hat{()}$, the total stress, strain and displacement fields are expressed as
\begin{equation}
\label{eq:superposition}
\bm{\sigma}=\hat{\bm{\sigma}}+\tilde{\bm{\sigma}},\quad \bm{\varepsilon}=\hat{\bm{\varepsilon}}+\tilde{\bm{\varepsilon}},\quad \bm{u}=\hat{\bm{u}}+\tilde{\bm{u}},
\end{equation}
respectively. The following procedure \citep{Jaafar2008} is adopted to evaluate the image stress field. First, the elastic stress field due to dislocations in an infinite body, $\tilde{\bm{\sigma}}$, is obtained using \autoref{eq:disstress}; the tractions $\tilde{\bm{T}}=\tilde{\bm{\sigma}}\cdot\bm{n}$ on the traction boundaries due to this stress are then calculated, reversed and added to the prescribed traction boundary conditions $\bm{T}$; these modified boundary values, $\hat{\bm{T}} = \bm{T}-\tilde{\bm{T}}$, in addition to any prescribed displacement conditions $\bm{U}$ on the displacement boundaries are used in an elastic finite element simulation to determine the corrective fields. Finally, the corrective stress field, $\hat{\bm{\sigma}}$, is used to evaluate the corrective nodal force $\hat{\bm{f}}$.

\subsection{Hydrogen elastic shielding}

The hydrogen stress contribution required for three dimensional DDD process was recently obtained by \citet{Gu2018} using isotropic linear elasticity theory, based on the work of \citet{Cai2014}. Hydrogen atoms were viewed as point defects of Eshelby type \citep{Eshelby1957}, and it was assumed that hydrogen diffusion is sufficiently rapid that hydrogen atoms can move with the dislocations. Therefore equilibrium of chemical potential is guaranteed over the domain. This was referred to as the \lq\lq fast diffusion\rq\rq $\ $scenario which is suitable for bcc materials. Under this assumption, steady-state hydrogen redistribution is established at all times, and is only a function of the dislocation pressure field $\tilde{P} = -\mathrm{Tr}(\tilde{\bm{\sigma}})/3$,
\begin{equation}
\label{eq:Hdistribution}
\chi(\bm{x})=\left[1+\frac{1-\chi_0}{\chi_0}\mathrm{exp}\left(\frac{\tilde{P}(\bm{x})\Delta V}{k_BT}\right)\right]^{-1},
\end{equation}
where $0 \leq \chi(\bm{x})\leq 1$ is the hydrogen field distribution expressed as the occupancy. Occupancy is related to hydrogen concentration $c(\bm{x})$ through $c(\bm{x})=\chi(\bm{x})c_{max}$, where $c_{max}$ is the maximum concentration or equivalently, the number of atomic sites per unit volume where each site can accomodate one hydrogen atom. The volume expansion of the matrix due to a hydrogen atom is $\Delta V$, and $\chi_0$ is the reference occupancy under a uniform pressure $P$
\begin{equation}
\label{eq:Hdistribution0}
\chi_0=\left[1+\mathrm{exp}\left(\frac{\mu_0+P\Delta V}{k_BT}\right)\right]^{-1},
\end{equation}
with $\mu_0$ being the reference chemical potential of hydrogen in a stress-free infinite medium. In reality, $\mu_0$ could represent the chemical potential of hydrogen in a remote reservoir and should be viewed as the initial hydrogen potential in the bulk.

The hydrogen population as distributed point defects will generate an elastic stress field which is dependent on the spatial distribution of hydrogen occupancy; which is a function of stress. Therefore the hydrogen induced stress is dependent on the dislocation configuration in the domain, as indicated in \autoref{eq:Hdistribution}. According to \citet{Gu2018}, the hydrogen stress field in three dimensions can be expressed as
\begin{align}
\label{eq:Hstress}
\bar{\boldsymbol{\sigma}}^{12}(\bm{x}) =&
-\frac{G^2(1+\nu)^2\Delta V^2}{18\pi (1-\nu)^2k_BT}c_{max}\chi_0(1-\chi_0)
\left[\int_{\bm{X}_1^H}^{\bm{X}_2^H}\left(\frac{1}{R_a^3}+\frac{3a^2}{R_a^5}\right)\left[(\bm{R}\times\bm{b}_{12})\cdot \mathrm{d}\bm{x'}\right]\cdot\bm{I}\right.\\
&-\int_{\bm{X}_1^H}^{\bm{X}_2^H}\frac{1}{R_a^3}\left[(\bm{b}_{12}\times \mathrm{d}\bm{x'})\otimes\bm{R}+\bm{R}\otimes(\bm{b}_{12}\times \mathrm{d}\bm{x'})\right]
\left.+\int_{\bm{X}_1^H}^{\bm{X}_2^H}\frac{3}{R_a^5}\left[(\bm{R}\times\bm{b}_{12})\cdot \mathrm{d}\bm{x'}\bm{R}\otimes\bm{R}\right] \right]. \nonumber
\end{align}
This is proportional to the last three terms (the none screw terms) of the dislocation stress field in \autoref{eq:disstress}, indicating that the hydrogen elastic stress field can be evaluated via a similar approach to evaluating the stress generated by a dislocation segment. Following \citet{Gu2018}, the hydrogen induced PK force $\bar{\bm{f}}$ is based on virtual hydrogen-related dislocation segments as illustrated in \autoref{fig:segs}(b). Due to the fast diffusion assumption for bcc materials, the hydrogen-related dislocation segments are simply the dislocation segments, and the hydrogen induced force at a node at $\bm{X}_3$ on the dislocation segment $4\rightarrow 3$ arising from the hydrogen stress field $\bar{\boldsymbol{\sigma}}^{12}$ is calculated using
%
\begin{align}
\label{eq:pkdisseg}
\bar{\bm{F}}_{3}=\bar{\bm{f}}^{12}_{43}(\bm{X}_3) & = \int_0^{L_{43}}\frac{l}{L_{43}}\left(\bar{\bm{\sigma}}^{12}\left[\bm{x}(l)\right]\cdot\bm{b}_{43}\right)\times\bm{l}_{43}\mathrm{d}l\\
\textrm{where}\quad\bm{x}(l)&=\left(1-\frac{l}{L_{43}}\right)\bm{X}_4+\frac{l}{L_{43}}\bm{X}_3, \nonumber
\end{align}
and this term needs to be added to the right hand side of \autoref{eq:Fk}.

In a finite volume, the hydrogen stress field $\bar{\bm{\sigma}}$ generates tractions on the surfaces of the body. This will cause the body to deform slightly inducing an additional image stress which will contribute to the force on the dislocation nodes. This corrective stress field $\hat{\bm{\sigma}}$ which includes the additional hydrogen image stress is obtained using FEM with the modified traction boundary condition $\hat{\bm{T}} = \bm{T} - \tilde{\bm{T}} - \bar{\bm{T}}$. The total elastic fields are then the superposition of the corrective FEM fields, $\hat{(.)}$, the analytic dislocation fields $\tilde{(.)}$ and analytic hydrogen fields $\bar{(.)}$
\begin{equation}
\label{eq:superpositionH}
\bm{\sigma}=\hat{\bm{\sigma}}+\tilde{\bm{\sigma}}+\bar{\bm{\sigma}},\quad \bm{\varepsilon}=\hat{\bm{\varepsilon}}+\tilde{\bm{\varepsilon}}+\bar{\bm{\varepsilon}},\quad \bm{u}=\hat{\bm{u}}+\tilde{\bm{u}}+\bar{\bm{u}}.
\end{equation}
After obtaining $\hat{\bm{\sigma}}$ and consequently $\hat{\bm{f}}$, the corrective force which includes the hydrogen image force, the total nodal force on node $k$ can be evaluated
\begin{equation}
\label{eq:FkH}
\bm{F}_{k} = \sum_{l}\sum_{i,j}\left(\tilde{\bm{f}}^{ij}_{kl}(\bm{X}_k)+\bar{\bm{f}}^{ij}_{kl}(\bm{X}_k)\right) + \hat{\bm{f}}(\bm{X}_k) + \bm{f}^c(\bm{X}_k),
\end{equation}
where $\bar{\bm{f}}^{ij}_{kl}$ is the elastic force due to the hydrogen segment $i\rightarrow j$ integrated along segment $k\rightarrow l$.

\subsection{Hydrogen dependent mobility law}
\label{Hmobility}

\begin{figure}[!h]
 \centering\includegraphics[width=0.9\linewidth]{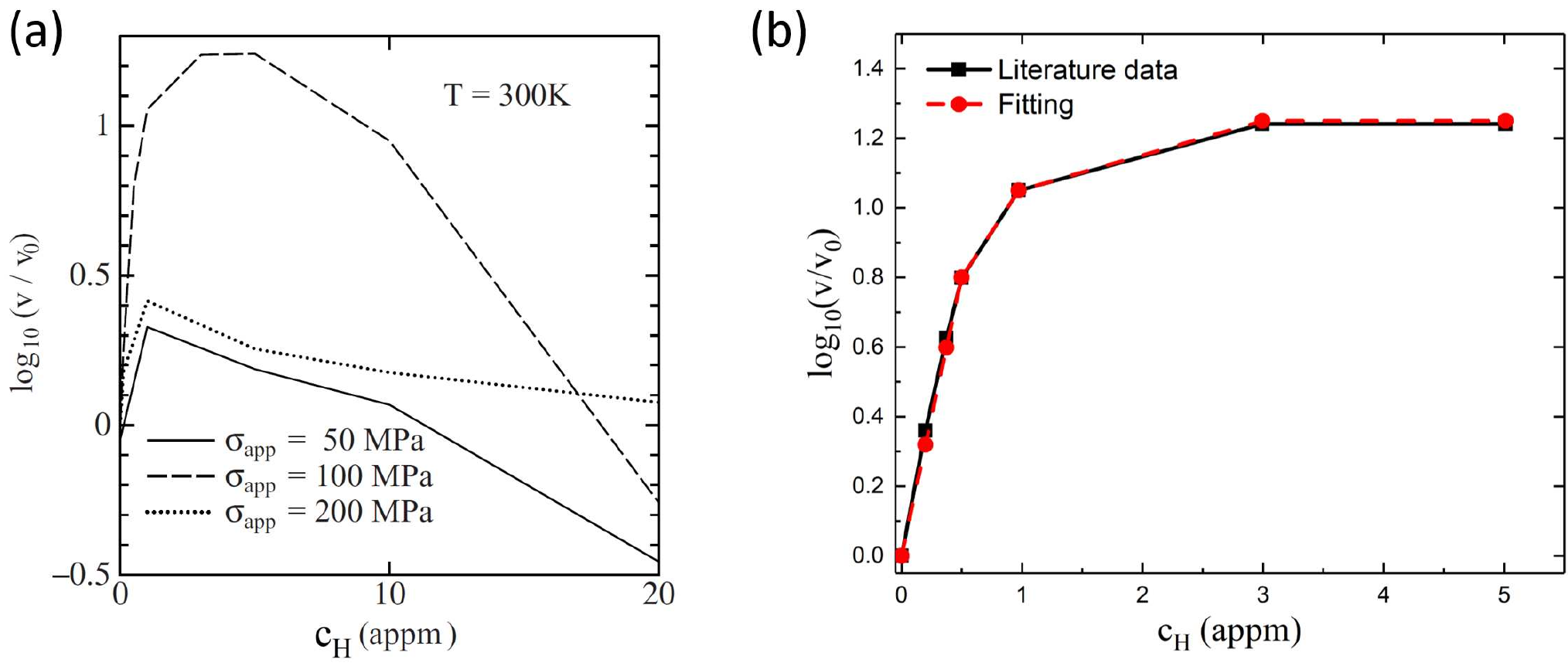}
\caption{(a) The effect of Hydrogen on screw mobility, where the mobility is represented by the logarithm of the dislocation velocity with hydrogen normalised by the velocity without hydrogen. This figure is reproduced from \citet{Katzarov2017} with permission; (b) hydrogen enhanced mobility law implemented in the current DDD and DDP simulations.}
\label{fig:Hmobility}
\end{figure}

The influence of hydrogen on dislocation mobility can be directly captured by atomistic simulations. \citet{Katzarov2017} calibrated the effect of hydrogen on the mobility of pure screw dislocations in $\alpha$-Fe, using kMC based on first principle calculations  \citep{Itakura2013}. A quantitative relationship between dislocation mobility and bulk hydrogen concentration were extracted under different temperatures and applied stresses. This work concerns the ambient temperature case, and the corresponding curves is reproduced with permission from \citet{Katzarov2017} in \autoref{fig:Hmobility}(a). The hydrogen effects are evaluated under three applied stress levels, and hydrogen enhanced dislocation mobility was observed up to $10$ appm in all the cases.

Hydrogen effects on dislocation mobility are incorporated into the DD simulation based on \autoref{fig:Hmobility}(a). In order to maintain a linear mobility law, the drag coefficients are assumed to be dependent on hydrogen but independent of applied stress. The curve corresponding to $\sigma_{\mathrm{app}}=100$ MPa is adopted throughout. Further, considering the small bulk hydrogen concentration in bcc materials, only the initial part of the curve $0.0\leq c_H \leq 5.0$ appm was fitted. Plotting this part in \autoref{fig:Hmobility}(b), the following linear fitting was obtained,
\begin{equation}
\label{eq:Hmobilityfit}
\lambda(c_H) = \log_{10}(\textrm{v}_H/\textrm{v})=\left\{ \begin{array}{ll}
1.6c_H,\qquad\quad 0<c_H\leq0.5 \ \mathrm{appm}\\
0.5c_H+0.55, \quad 0.5<c_H\leq1.0 \ \mathrm{appm}\\
0.1c_H+0.95, \quad 1<c_H\leq3.0 \ \mathrm{appm}\\
1.25,\qquad\quad c_H>3.0 \ \mathrm{appm}
\end{array} \right..
\end{equation}
According to this equation, the increase in dislocation mobility due to hydrogen can be as large as $17$ times with $c_H\ge3$ appm. In this work, however, the maximum increase achieved is less than $10$ due to the small initial bulk concentration used.

In DD simulations, the local hydrogen concentration $c_H^{ij}$ at each segment is evaluated during each iteration. The drag coefficient for each segment is  scaled using \autoref{eq:Hmobilityfit} and the nodal velocity is then found using \autoref{eq:linearmob}. The scalar drag coefficients used to construct the drag tensor \autoref{eq:Bkl} for each segment were reduced using
\begin{equation}
\label{eq:Hdrag}
B_H = 10^{-\lambda(c_H)}B,
\end{equation}
where $B$ is the hydrogen free drag coefficient, either $B_{s}$ or $B_{eg}$. Note that $B_{ec}$ was unchanged to ensure climb did not occur.

The local hydrogen concentration $c_H(\bm{x})$ at a field point $\bm{x}$ is obtained using \autoref{eq:Hdistribution}. However, the normal components of the non-singular stress field of a straight dislocation segment are all zero on the dislocation line and so $\textrm{Tr}(\bm{\sigma}) = 0$ when the field point is on a dislocation segment. Therefore, in order to evaluate the concentration on a dislocation segment without neglecting the self stress of the segment, the concentration was found at 2 field points at $\pm a\bm{n}$, either side of the segment midpoint, and then averaged. Where $a$ is the core width in \autoref{eq:disstress} and $\bm{n}$ the glide plane normal of the segment.

The kMC investigation into hydrogen effects on screw  mobility in \citet{Katzarov2017} was performed considering hydrogen concentration near the dislocation core which was well captured by first principle calculations \citep{Itakura2013}. Starting from a sensible bulk hydrogen concentration, e.g. $c_H=10$ appm a very high hydrogen occupancy up to $\chi=0.05$ (equivalent to $c=5\times10^3c_H$) could build up in the core region due to its high binding energy. Such a high concentration is impossible to achieve within linear elasticity theory. This implies the hydrogen increased dislocation mobility could be viewed as the total effect including the contribution of hydrogen in the core region, whereas hydrogen elastic shielding considers only the weak hydrogen accumulation generated by the remote dislocation elastic stress fields. The local hydrogen concentration obtained in the current DDD simulation should be considered equivalent to the bulk concentration $c_H$ \citep{Katzarov2017} instead of the equilibrium concentration at binding sites in atomic simulations.

Exactly how hydrogen influences edge mobility is unclear, since no quantitative description is available and there is consensus in the literature. \cite{Robertson2001} concluded that hydrogen increases dislocation mobility equally for edge and screw dislocations, while atomistic simulations predict hydrogen reduces edge mobility \citep{Song2014,Bhatia2014} and increases screw mobility \citep{Itakura2013,Katzarov2017}. Experimentally, \citet{Xie2016} observed hydrogen decreased edge dislocation mobility in Al during in situ micropillar cyclic loading test, and the same author \citep{Xie2018} most recently observed hydrogen enhanced screw mobility in Fe using similar technique. Therefore we tested two cases. Case $I$ assumes hydrogen only increases screw mobility, and case $II$ assumes an equal influence on screw and edge mobility
\begin{equation}
\label{eq:cases}
\begin{aligned}
&\mathrm{Case}\ I: B_{s,H} = 10^{-\lambda(c_H)}B_s \quad \textrm{and} \quad B_{eg,H} = B_{eg},\\
&\mathrm{Case}\ II: B_{s,H} = 10^{-\lambda(c_H)}B_s \quad \textrm{and} \quad B_{eg,H} = 10^{-\lambda(c_H)}B_{eg},
\end{aligned}
\end{equation}
with $\lambda_H(c_H)$ defined in \autoref{eq:Hdrag} where the concentration field varies in space $c_H= c_H(\bm{x})$ and is evaluated at the segment midpoint.  For cases $I$, it is noted that screw segments still possess lower mobility than edge segments in the presence of hydrogen in all simulations as $B_s=20B_{eg}$.

\section{Hydrogen effects in an infinite medium}
\label{infinite}


\subsection{Geometry for hydrogen informed DDD simulation}
\label{DDDmodel}
With hydrogen informed DDD simulation, the influence of hydrogen on dislocation generation, pile-up and slip planarity is investigated starting from a FR source in an infinite solid subjected to both zero and non-zero out-of-plane PK forces. The findings here will provide rationalization for the phenomena observed in microcantilever bending simulations in section \ref{sec:finite}. In all the simulations, two different hydrogen effects, hydrogen elastic shielding and hydrogen increased dislocation mobility, are considered and their contributions evaluated. Subsequently, the model geometry and loading conditions are introduced first, and the selection of model parameters is then elaborated.

\begin{figure}[!h]
\centering\includegraphics[width=0.4\linewidth]{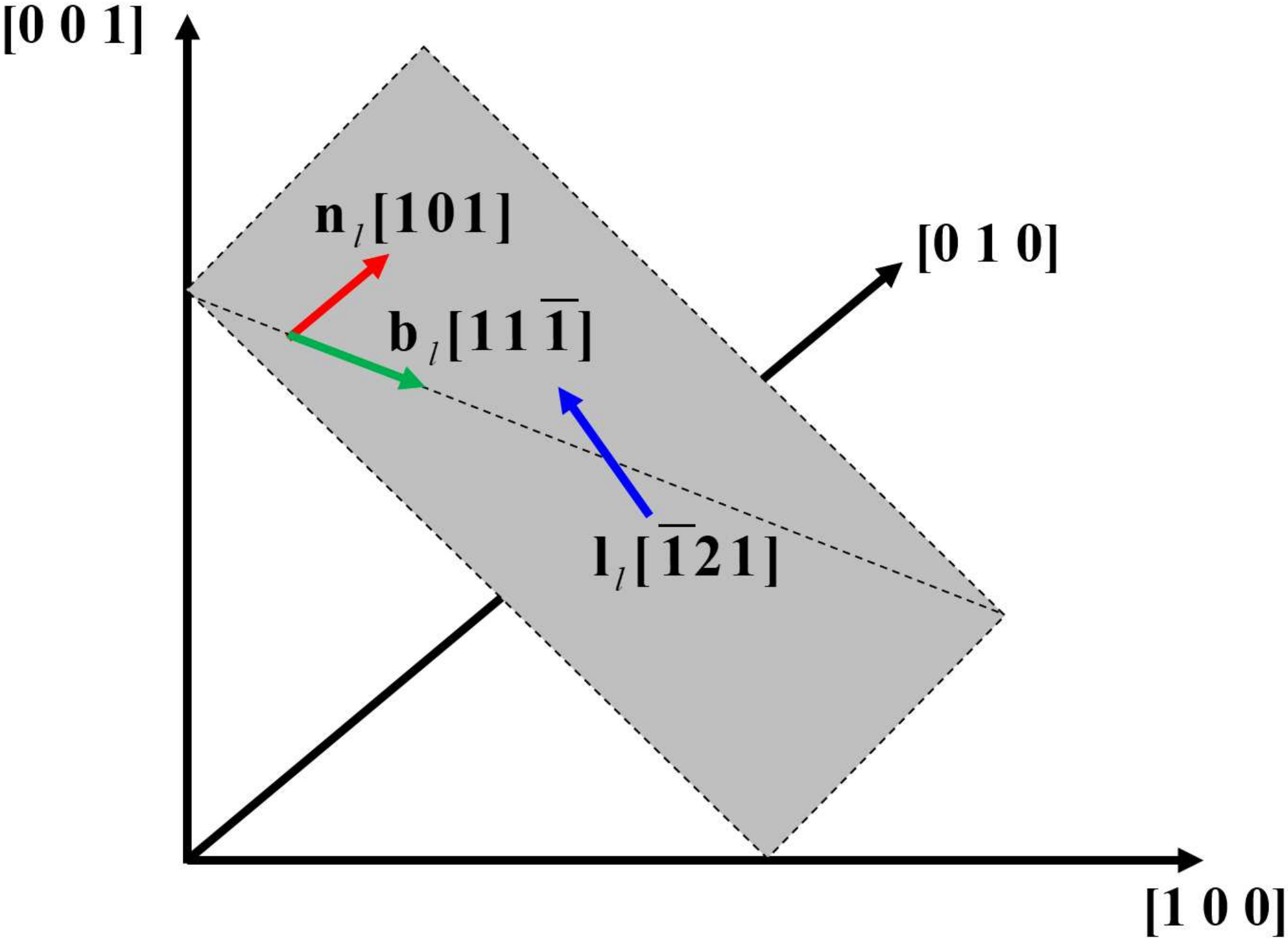}
\caption{Illustration of a pure edge dislocation belonging to the $(101)[11\bar{1}]$ slip system in local crystallographic coordinate system. The edge dislocation is pinned at both ends, and its line direction is $\bm{l}_l=(-1,2,1)$. The vectors in this local coordinate system are marked with a subscript $_l$.}
\label{fig:localsys}
\end{figure}

\begin{figure}[!h]
\centering\includegraphics[width=0.6\linewidth]{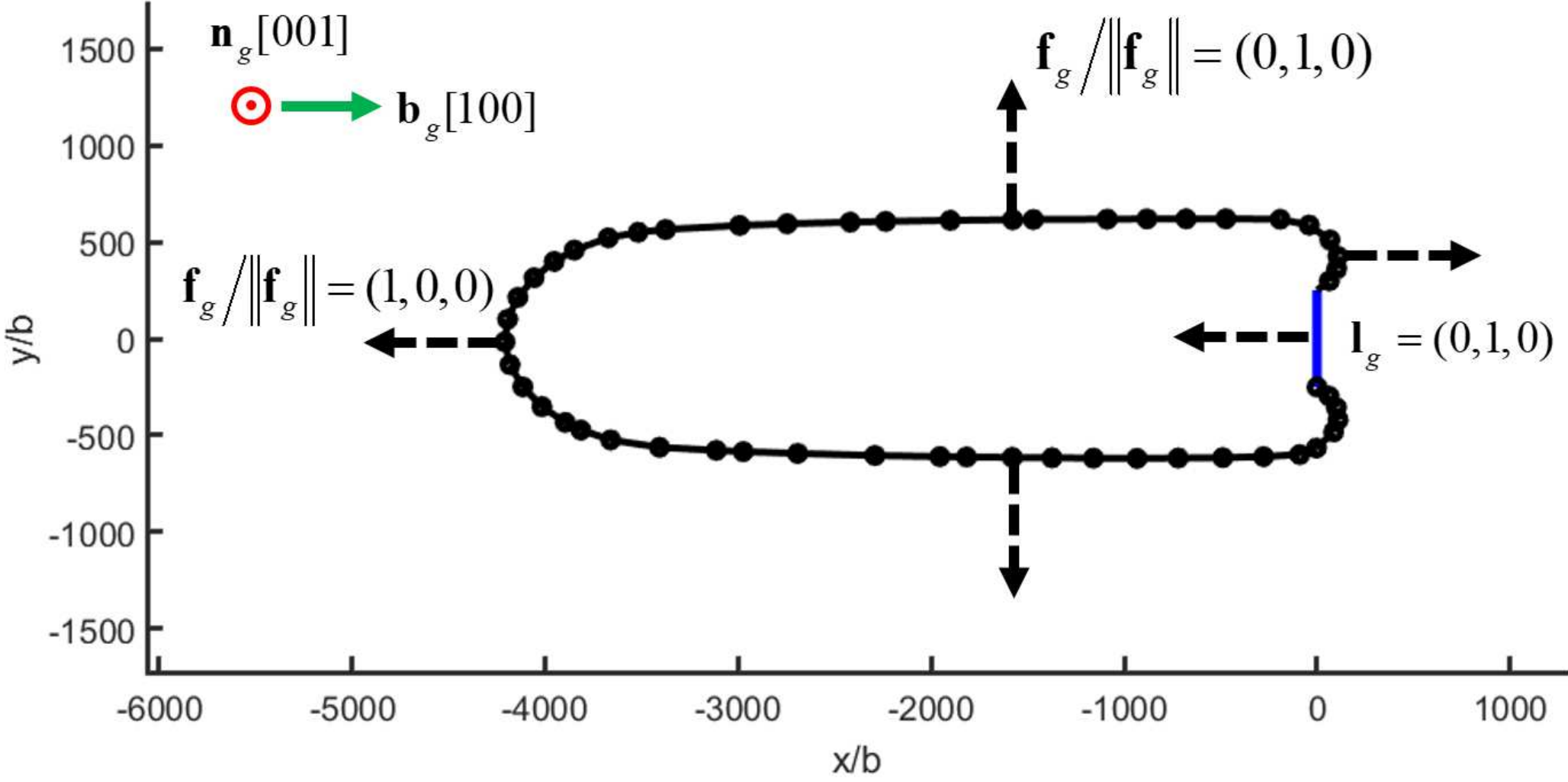}
\caption{Illustration of the FR source in the global coordinate system, subjected to simple shear in \autoref{eq:globalstress1}. The initial configuration is marked blue, the dislocation nodes also plotted. The direction of the PK forces acting on pure screw and pure edge segments are indicated with dashed arrows.}
\label{fig:FRglobal}
\end{figure}

\begin{figure}[!h]
\centering\includegraphics[width=0.6\linewidth]{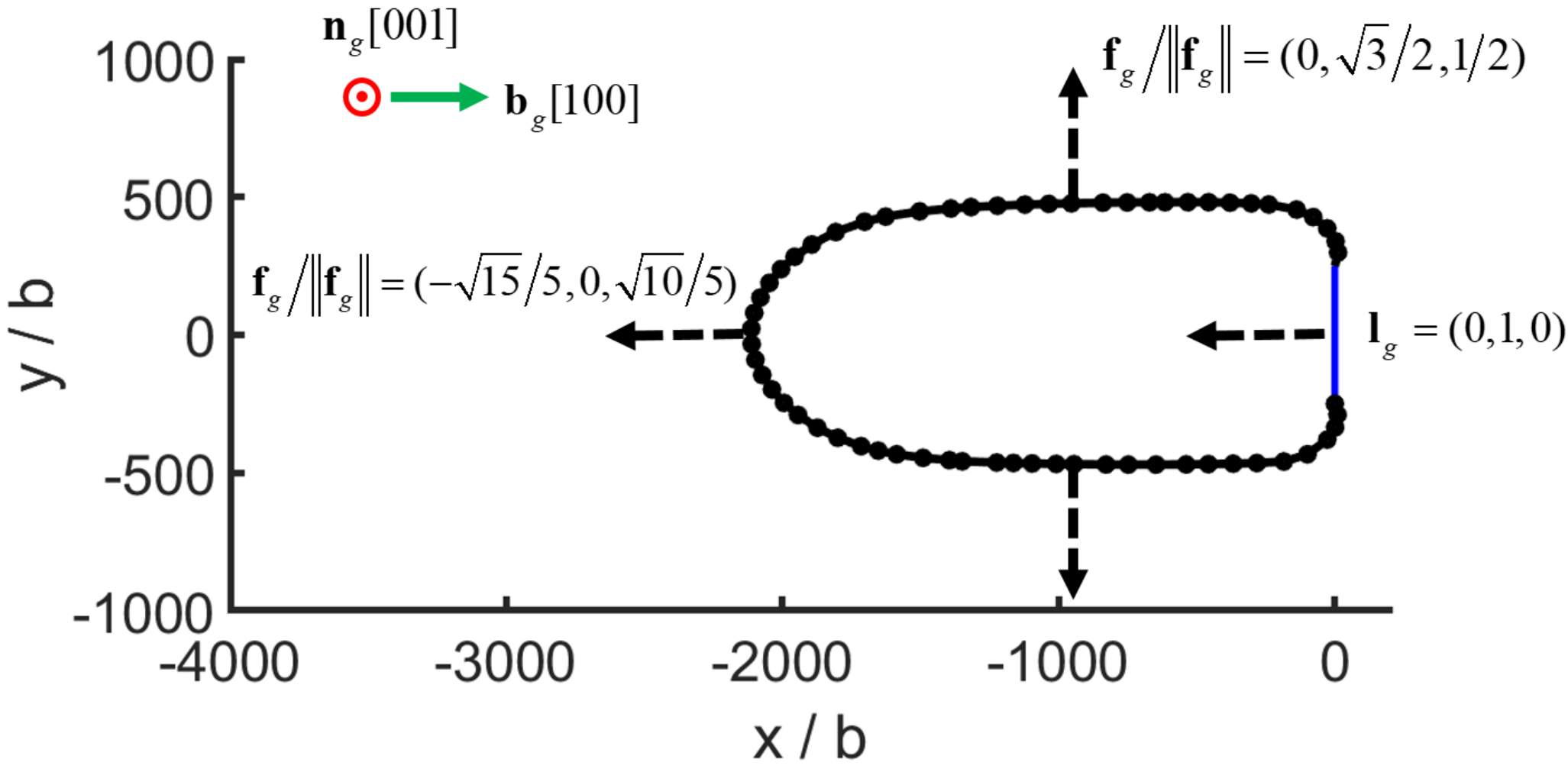}
\caption{Illustration of out of plane forces in the global (dislocation) coordinate system, resulting from the stress state in \autoref{eq:globalstress2}. The initial configuration is marked blue, dislocation segments and nodes at a later time are also plotted. The direction of the PK forces arising from the applied stress are indicated with dashed arrows.}
\label{fig:planarityglobal}
\end{figure}

A pure edge dislocation belonging to the $(101)[11\bar{1}]$ slip system is selected as the initial configuration for DDD simulation in an infinite medium. The dislocation lies on the $(101)$ plane and is pinned at both ends, as shown in \autoref{fig:localsys}. Its behavior is dependent on the applied stress $\boldsymbol{\sigma}_l$ in the local coordinate system. For easier illustration, the model is rotated into a global system, with axis along $\bm{b}_l$, $\bm{l}_l$ and $\bm{n}_l$, as shown in \autoref{fig:FRglobal} and \autoref{fig:planarityglobal}. The stress tensors in the global coordinate system are readily obtained
\begin{equation}
\label{eq:stressconvert}
\bm{\sigma}_{g}=\bm{R}\bm{\sigma}_{l}\bm{R}^T
\end{equation}
where $\bm{R}$ is the rotation matrix from the local (crystal) to the global (dislocation) frame
\begin{equation}
\label{eq:rotation}
\bm{R}=
\left[ {\begin{array}{c}
   \bm{b}_l/\lvert\bm{b}_l\rvert \\
   \bm{l}_l/\lvert\bm{l}_l\rvert \\
   \bm{n}_l/\lvert\bm{n}_l\rvert
  \end{array} } \right]
  =
\left[ {\begin{array}{ccc}
   1/\sqrt{3} & 1/\sqrt{3} & -1/\sqrt{3}\\
   -1/\sqrt{6} & 2/\sqrt{6} & 1/\sqrt{6}\\
   1/\sqrt{2} & 0 & 1/\sqrt{2}
  \end{array} } \right].
\end{equation}

In \autoref{fig:FRglobal}, out-of-plane pure shear is applied
\begin{equation}
\label{eq:globalstress1}
\bm{\sigma}_{g}=
\left[ {\begin{array}{ccc}
   0 & 0 & \tau\\
   0 & 0 & 0\\
   \tau & 0 & 0
  \end{array} } \right].
\end{equation}
The PK forces on all dislocation components are in plane, therefore, the initial edge dislocation behaves as an FR source.
In \autoref{fig:planarityglobal}, uniaxial tension along the crystallographic $[100]$ direction in the local coordinate system is applied, which in the global system is expressed as
\begin{equation}
\label{eq:globalstress2}
\bm{\sigma}_{g}=
\bm{R}\left[ {\begin{array}{ccc}
   \sigma & 0 & 0\\
   0 & 0 & 0\\
   0 & 0 & 0
  \end{array} } \right]\bm{R}^T.
\end{equation}
At the beginning, the pure edge dislocation bows out to the left driven by the PK force with a component in the negative x direction. Although this PK force possesses a non-zero component in the positive z direction, it does not cause any out-of-plane motion of the edge segment, since climb is not permitted. During the bowing out, pure screw segments are generated, which move in the y direction and cross slip. Therefore, when the dislocation bows back, the screw segments are unable to annihilate each other unless they first cross slip back onto the same plane. The simple model illustrated in \autoref{fig:FRglobal} is used to investigate hydrogen effects on dislocation generation and pile-up, and that in \autoref{fig:planarityglobal} is used to study hydrogen effects on slip planarity, with the results elaborated in \autoref{FRresults} and \autoref{Planarityresults}, respectively. 

Adopting the material parameters (relevant to Ni), initial loop size and hydrogen concentration values used by \citet{Gu2018}, the same results in terms of PK force magnitude and loop shape change were obtained, which are not repeated here. With the current implementation validated, material parameters relevant to Fe, were used to simulate an FR source. The Burgers vector magnitude is $\lvert\bm{b}\rvert=\frac{\sqrt{3}}{2}a_0$, where $a_0=2.856$ \r{A} is the lattice parameter; isotropic linear elasticity is assumed, with a shear modulus $\mu=83$ GPa and Poisson's ratio $\nu=0.29$ \citep{Tang2014}. In the absence of hydrogen, the edge and screw drag coefficients are $B_{eg}=5\times10^{-4}~\textrm{Pa s}$ and $B_{s}=1\times10^{-2}~\textrm{Pa s}$ as used by \citet{Wang2011}. Climb is disabled in the current work, therefore, only pure screw segments can possibly move out of the glide plane, and a dislocation segment is considered pure screw if the angle $\theta$ between its line direction and Burgers vector is $\theta<0.01^\circ$. The dislocation core width is selected as $a=10 \mathrm{b}$, as used by \citet{Arsenlis2007} and \citet{Ayas2014}. For DDD simulation in an infinite medium (\autoref{fig:FRglobal} and \autoref{fig:planarityglobal}), the initial dislocation length is $l_0=500\mathrm{b}$ in both cases. The applied shear stress in \autoref{eq:globalstress1} is $\tau=450$MPa, and the uniaxial tension stress is $\sigma=800$MPa in \autoref{eq:globalstress2}. All simulations are for $T=300$ K, and the fast diffusion scenario \citep{Gu2018} is assumed, given the large diffusivity of hydrogen in bcc materials at this temperature. As for the initial bulk hydrogen concentration, an extremely large value $c_0=5\times10^5$ appm is selected to produce a noticeable effect with only hydrogen elastic shielding taken into account; a realistic value $c_0=0.1$ appm is then adopted in the simulations considering hydrogen increased dislocation mobility. 

\subsection{A Frank-Read source in the presence of hydrogen}
\label{FRresults}
\subsubsection{Hydrogen elastic shielding}
\label{FRshielding}

\begin{figure}[!h]
\centering\includegraphics[width=0.7\linewidth]{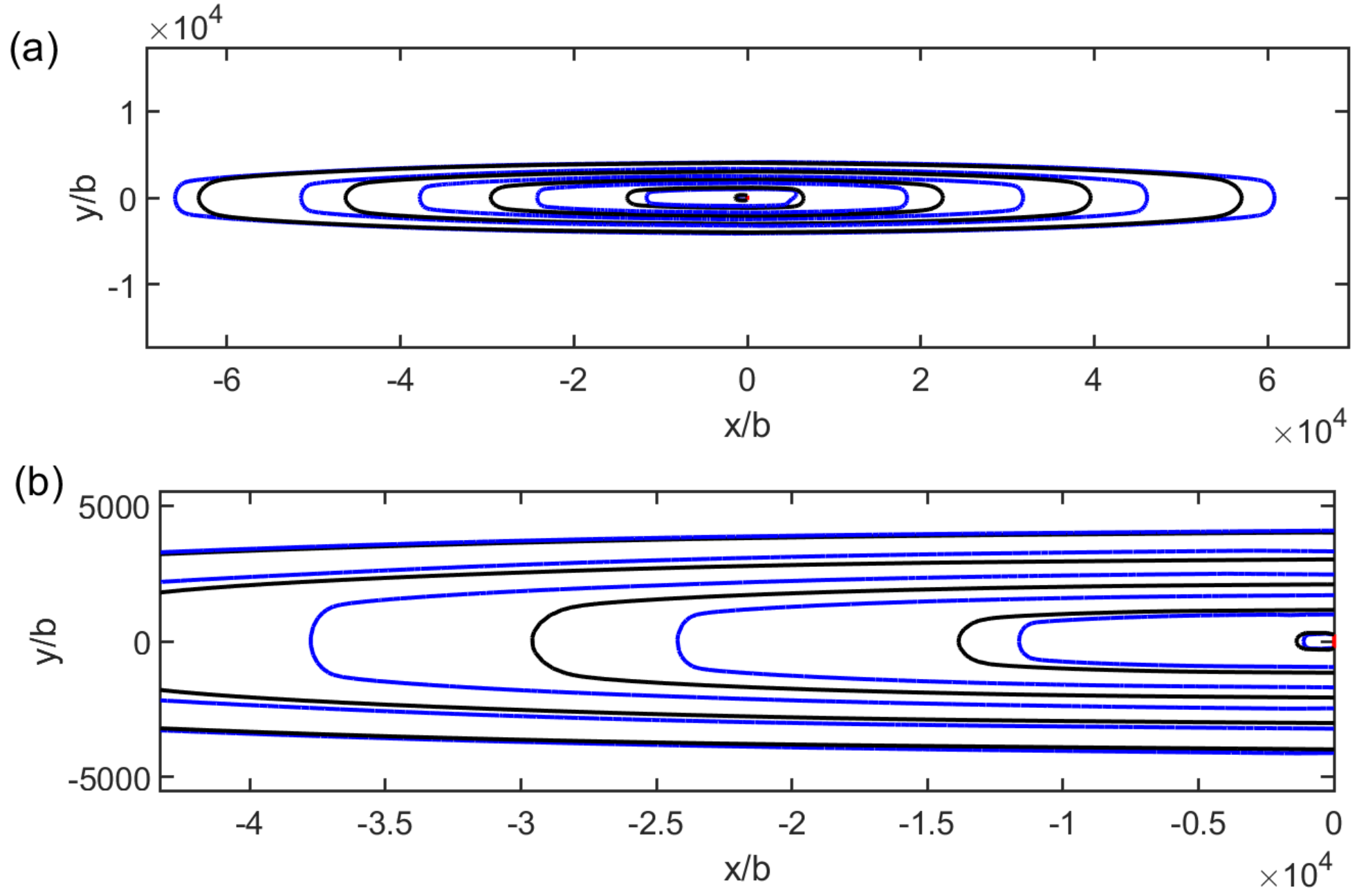}
\caption{(a) The black lines represent dislocation loops generated by an FR source in the absence of hydrogen, and the blue lines represent those generated with a hydrogen concentration of $c_0=5\times10^5$ appm. Both are plotted after the same simulated time, $t=87$ ns; the initial FR source is in red. (b)  Zooming-in on the central-left region of (a) shows an additional dislocation loop is generated in the presence of hydrogen.}
\label{fig:FRshielding}
\end{figure}

For Fe, the maximum hydrogen concentration $c_{max}$ in \autoref{eq:Hdistribution} is $\approx30\%$ lower than Ni. The lower mobility of screw components in Fe results in an elongated dislocation loop, as shown in \autoref{fig:FRshielding}, instead of a circular one in Ni. As the density of screw segments (which have no elastic interaction with hydrogen) is higher in Fe than in Ni the effect of hydrogen shielding is reduced for the same occupancy. Therefore, a higher hydrogen occupancy was required than by \cite{Gu2018}, in order to produce noticeable hydrogen influence.

Starting from an extremely high bulk concentration $c_0=5\times10^5$ appm, hydrogen elastic shielding influences the behavior of a FR source, as shown in \autoref{fig:FRshielding}. With hydrogen present the dislocation loops generated by the FR source move further, and one extra dislocation loop is generated, indicating that hydrogen shielding can promote dislocation nucleation. In addition, the average spacing between the dislocation loops is slightly decreased due to hydrogen, as also observed by \cite{Gu2018}. 

\begin{figure}[!h]
\centering\includegraphics[width=0.9\linewidth]{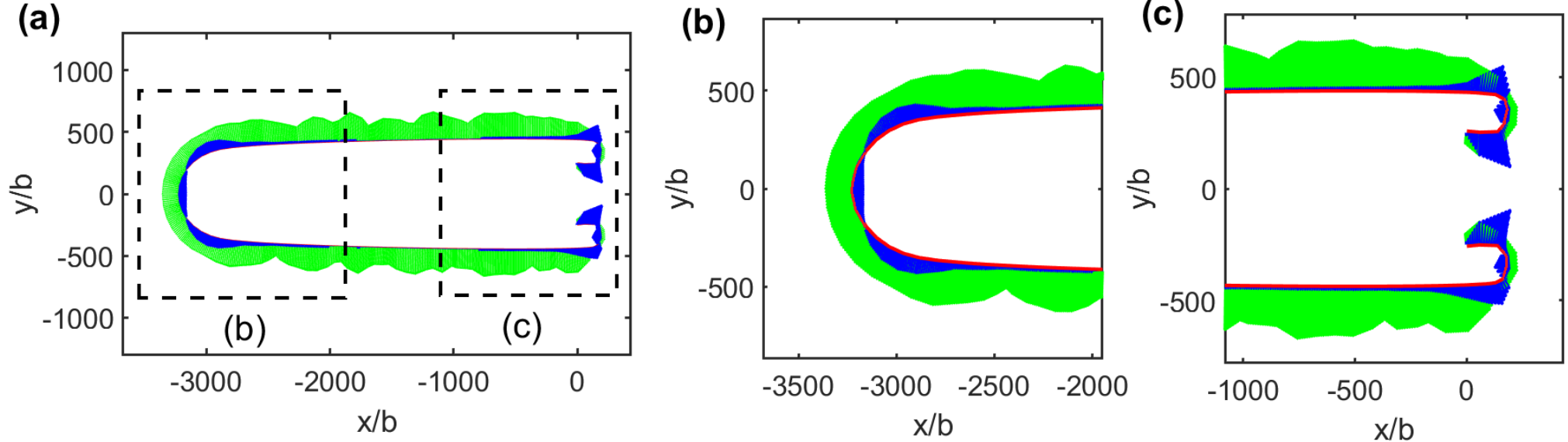}
\caption{ Partition of nodal forces on a dislocation half loop generated by the FR source prior to nucleation of the first loop. (a) The nodal force $\tilde{\bf{F}}_k+\hat{\bf{F}}$ due to the external and dislocation stress (green arrows), and hydrogen force $\bar{\bf{F}}_k$  (blue arrows) are shown acting on the dislocation nodes (red). $\bar{\bf{F}}_k$ arrows are scaled by $10\times$ in order to make them visible. (b) and (c) close up of regions indicatd in (a).}
\label{fig:FRshieldingforce2}
\end{figure}

The nodal force contribution due to the dislocation segments and the uniform applied stress $\tilde{\bf{F}}_k+\hat{\bf{F}}_k$ and due to the hydrogen stress $\bar{\bf{F}}_k$ (scaled $\times 10$) are plotted shortly after the first bow out of the FR source in \autoref{fig:FRshieldingforce2}(a). The direction of $\tilde{\bf{F}}_k+\hat{\bf{F}}$ tends to expand the half loop and cause it to pinch off to generate a full loop. The distribution of $\bar{\bf{F}}_k$ is highly dependent on the character of the segments connected to the node. In general, the values are always much less than the other elastic forces and are largest on the nodes connected to edge dominated segments. $\bar{\bf{F}}_k$ on nodes on long screw segments are negligible. This is expected as screw dislocations exert zero hydrostatic stress there is no elastic interaction with the hydrogen, e.g. using \autoref{eq:Hstress} shows $\bar{\bm{\sigma}}^{kl}=0$ for a pure screw segment. The hydrogen force, $\bar{\bf{F}}_k$, along pure edge segments or mixed segments which are edge dominant opposes the other elastic forces, $\tilde{\bf{F}}_k+\hat{\bf{F}}$, whereas the directions are the same for mixed segments which are screw dominant, as shown in \autoref{fig:FRshieldingforce2}(b). This indicates that hydrogen has a tendency to slow down the loop expansion along $\bf{b}$. As the anisotropy in the mobility is slightly reduced, this leads to a slightly more circular shape of the half loop.
In addition, \autoref{fig:FRshieldingforce2}(c) shows how the hydrogen force promotes the attraction of the short screw dominant segments. Making it easier for the segments to bow back and pinch off to nucleate a loop.
So far, hydrogen elastic shielding has been demonstrated to have a tendency to enhance dislocation nucleation and to decrease the average spacing. However, it should be noted that to achieve this required an extremely high bulk hydrogen concentration, e.g. $c_0=5\times10^5$ appm. With realistic bulk concentrations for bcc Fe, the elastic hydrogen effect becomes negligible. This was verified using $c_0=0.1$ appm. In this case, the FR source behaviour was identical to the hydrogen-free case in \autoref{fig:FRshielding} and so is not repeated.

\subsubsection{Hydrogen increased mobility}
\label{FRcaseI}
\begin{figure}[!h]
\centering\includegraphics[width=0.7\linewidth]{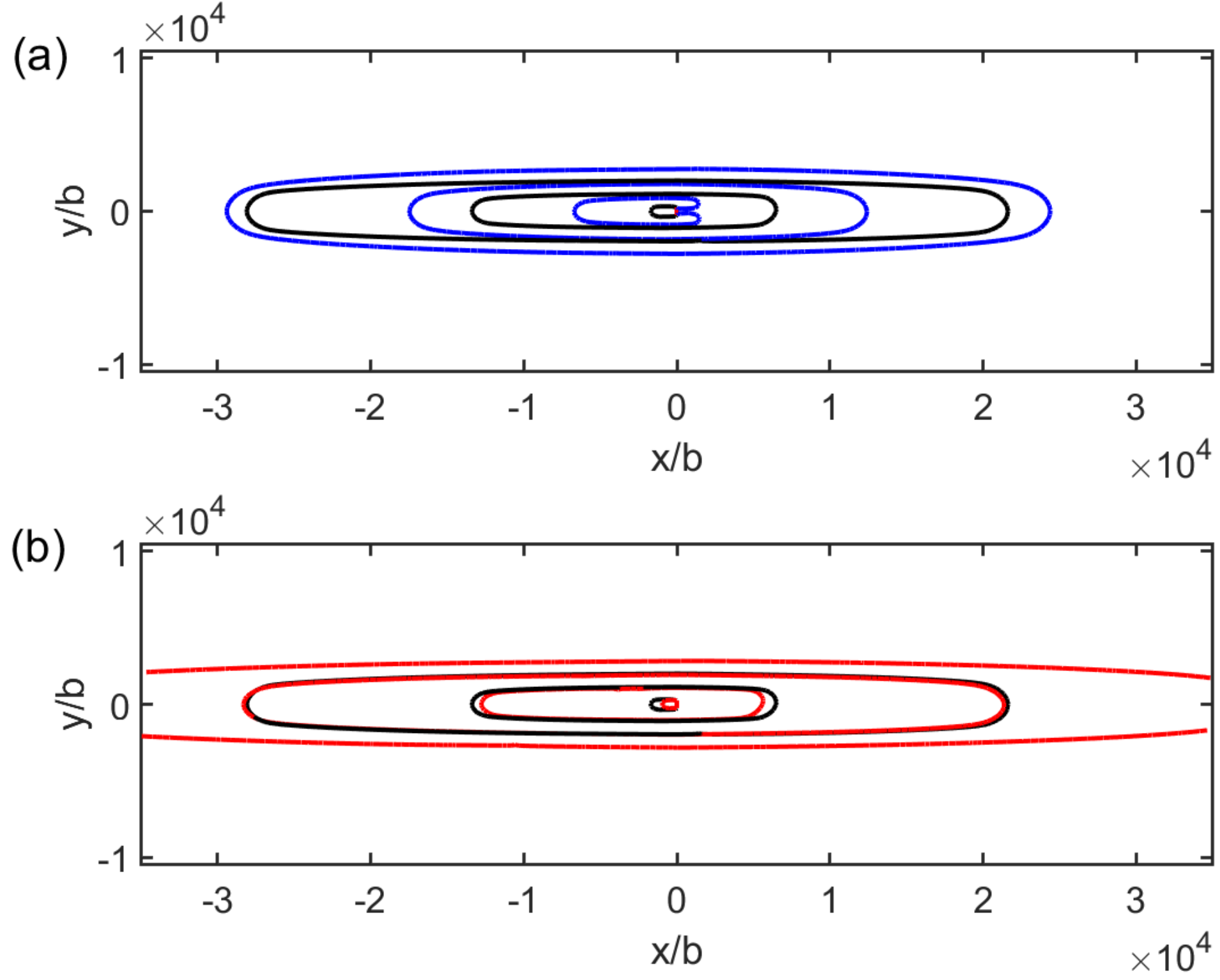}
\caption{Superimposed dislocation structures after the same simulated time $t=43$ ns without hydrogen (black lines) and with hydrogen at with a bulk concentration of $c_0=0.1$ appm. (a) shows case $I$, without hydrogen (black) and with hydrogen (blue); (b) shows case $II$, without hydrogen (black) and with hydrogen (red).}
\label{fig:FRmobilityobserve}
\end{figure}

\begin{figure}[!h]
\centering\includegraphics[width=0.7\linewidth]{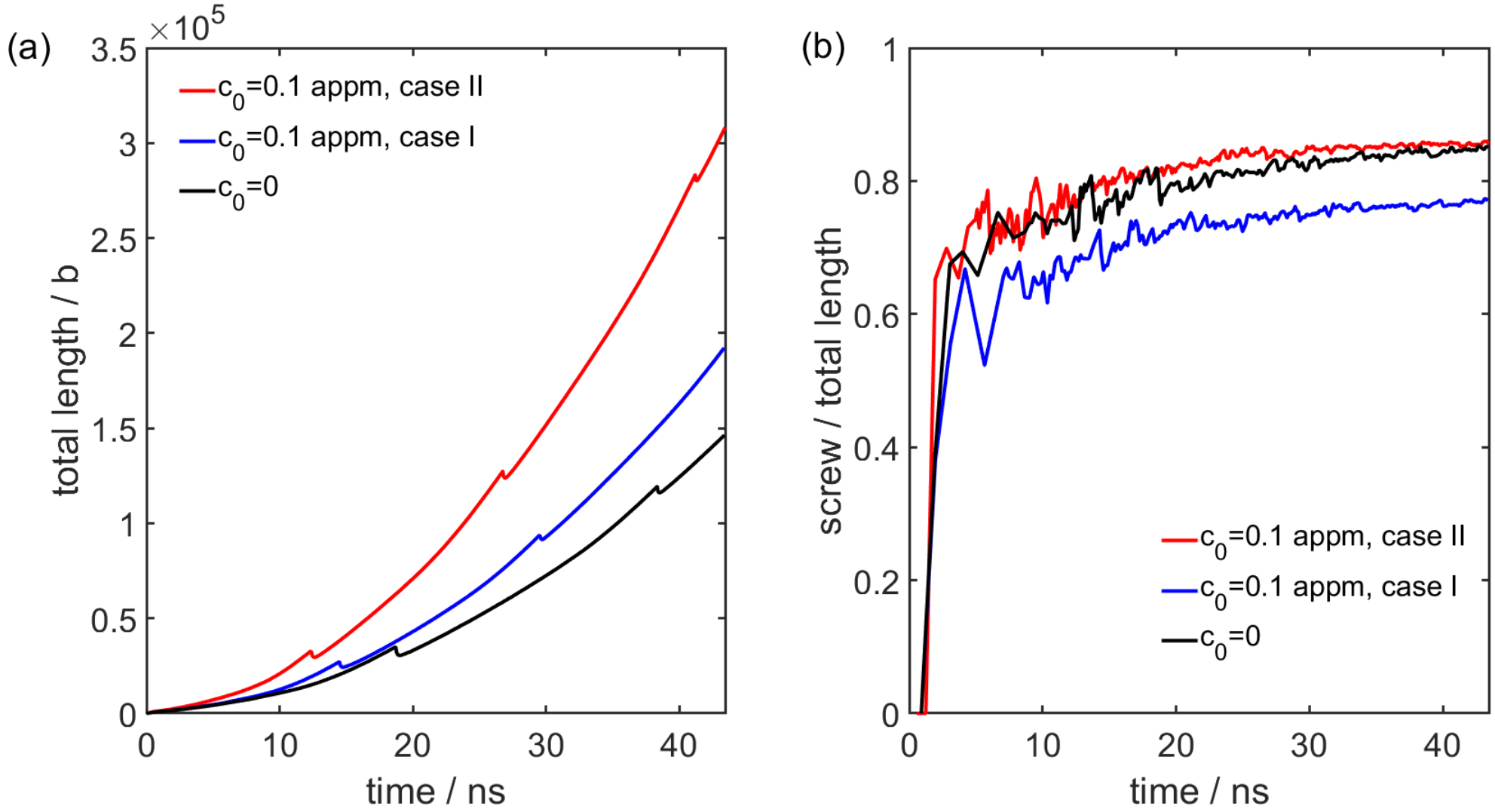}
\caption{(a) Evolution of the total dislocation line length with and without hydrogen; (b) evolution of the proportion of screw dominant segments with and without hydrogen.}
\label{fig:FRmobilitycurves}
\end{figure}

We now consider the effect of hydrogen increased mobility at a realistic bulk concentration of $c_0=0.1$ appm. Both case $I$ and case $II$, as outlined in \autoref{eq:cases}, are investigated. The dislocation structures with and without hydrogen after the same simulated time are superimposed in \autoref{fig:FRmobilityobserve}.
In both cases, more loops are generated, in the presence of hydrogen. The total dislocation line length increases with the hydrogen enhanced mobility as shown in \autoref{fig:FRmobilitycurves}(a). In case $I$, where hydrogen only enhances screw mobility, the shape of the loop differs from the hydrogen free situation. Whereas in case $II$, where hydrogen exerts equal influence on both screw and edge mobilities, the shape of the loops is unchanged.  \autoref{fig:FRmobilitycurves}(b) shows how the proportion of screw dominant segments evolves with simulated time. Here, a segment is considered screw dominant if the angle between $\bm{l}$ and $\bm{b}$ is less than $5^\circ$. It is noted this criterion is only used during postprocessing to identify screw like segments. The proportion of screw like segments is reduced in case $I$ but unchanged in case $II$. This same trend can also be reproduced be ignoring the spatial distribution of hydrogen completely and uniformly increasing the screw mobility (a simplified case I) or uniformly increasing the mobility of all segments (a simplifed case II). The same dislocation structure is predicted in case II as without hydrogen, but at an earlier simulated time. 

In case $I$, hydrogen promotes dislocation generation via increased screw mobility and the shape of loops becomes more circular; in case $II$, hydrogen accelerates dislocation evolution but has little influence on the final dislocation structure in quasi static loading. Case $I$  reduces the anisotropy in the dislocation motion, this was noted by \citet{Bhatia2014} and observed experimentally in \cite{Moriya1979}. Case $II$, however, finds no direct evidence in the literature except for a very general statement in \cite{Robertson2001}, besides, case II implies hydrogen does not alter the dislocation structure. This is inconsistent with experimental evidence on hydrogen enhanced plasticity. Therefore only case $I$, i.e. hydrogen increases screw mobility, is used in the following sections.

\subsection{Hydrogen effects on slip planarity}
\label{Planarityresults}

\subsubsection{Hydrogen elastic shielding}
\label{planarityshielding}
\citet{Ferreira1999} attempted to explain hydrogen enhanced slip planarity in terms of hydrogen elastic shielding, as reviewed in \autoref{intro:model}. From the energetic point of view, it was shown with linear elasticity theory that hydrogen elastic forces reduced the energy of edge dislocations relative to screw thereby reducing the frequency of cross-slip. It should be noted that this mechanism only occurred at a high concentration, $c_0>1\times10^4$ appm. Following this idea, we first investigate the role of hydrogen elastic shielding in enhanced slip planarity. 

\begin{figure}[!h]
\centering\includegraphics[width=0.9\linewidth]{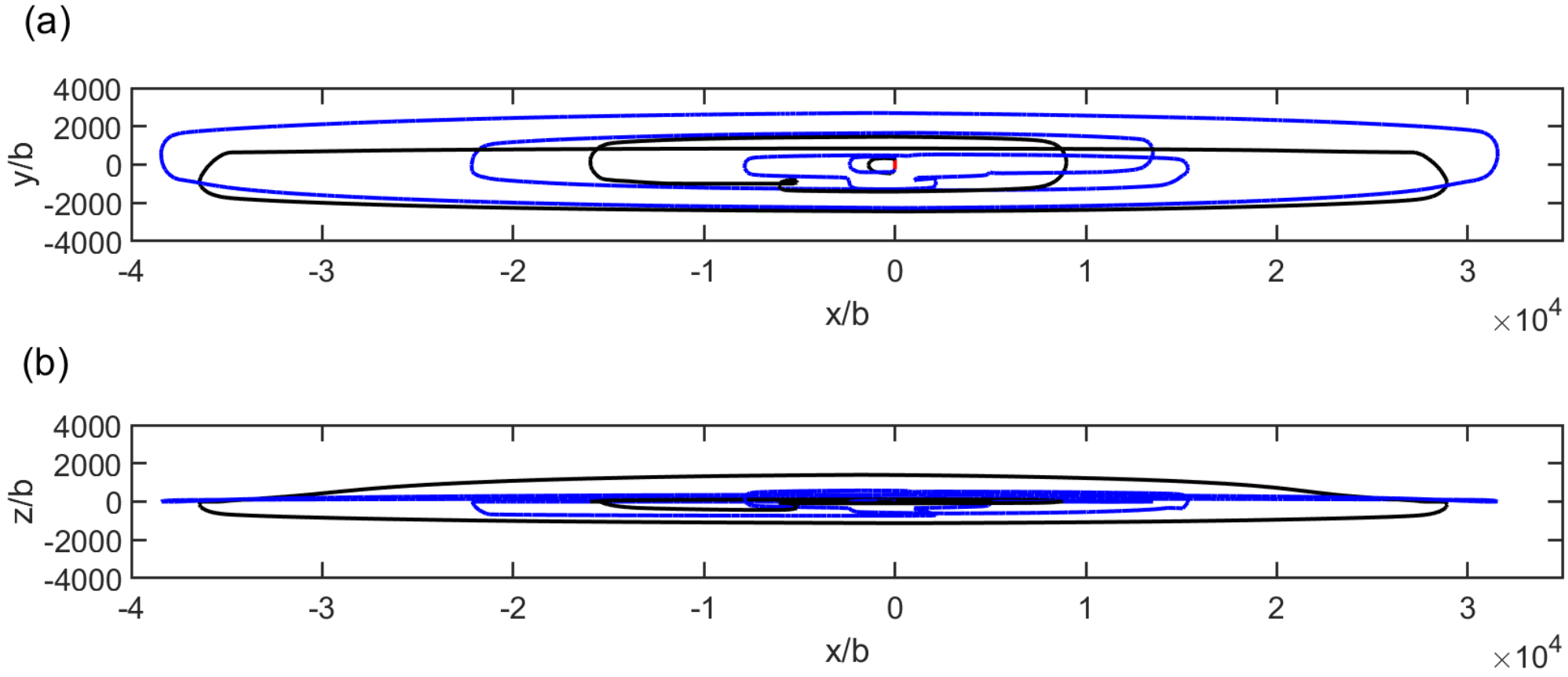}
\caption{Dislocation structures with and without hydrogen elastic shielding after the same simulated time, $t=80$ ns. (a) shows the projection of the dislocation structure on its slip plane, and (b) shows the out-of-plane motion along the plane normal. The dislocation structure without hydrogen is marked black, and that with hydrogen elastic shielding at a very high concentration $c_0=5\times 10^5$ appm is marked blue.}

\label{fig:planarityshieldingstruct}
\end{figure}

The applied stress was chosen to generate an out-of-plane PK force so that screw segments cross slip. Similar to the case in \autoref{FRshielding}, an extremely high bulk concentration $c_0=5\times10^5$ appm was used.  The dislocation structures with and without hydrogen are superimposed in \autoref{fig:planarityshieldingstruct}, and it is clear that out of plane motion is suppressed in the presence of hydrogen. 
%

\begin{figure}[!h]
\centering\includegraphics[width=0.6\linewidth]{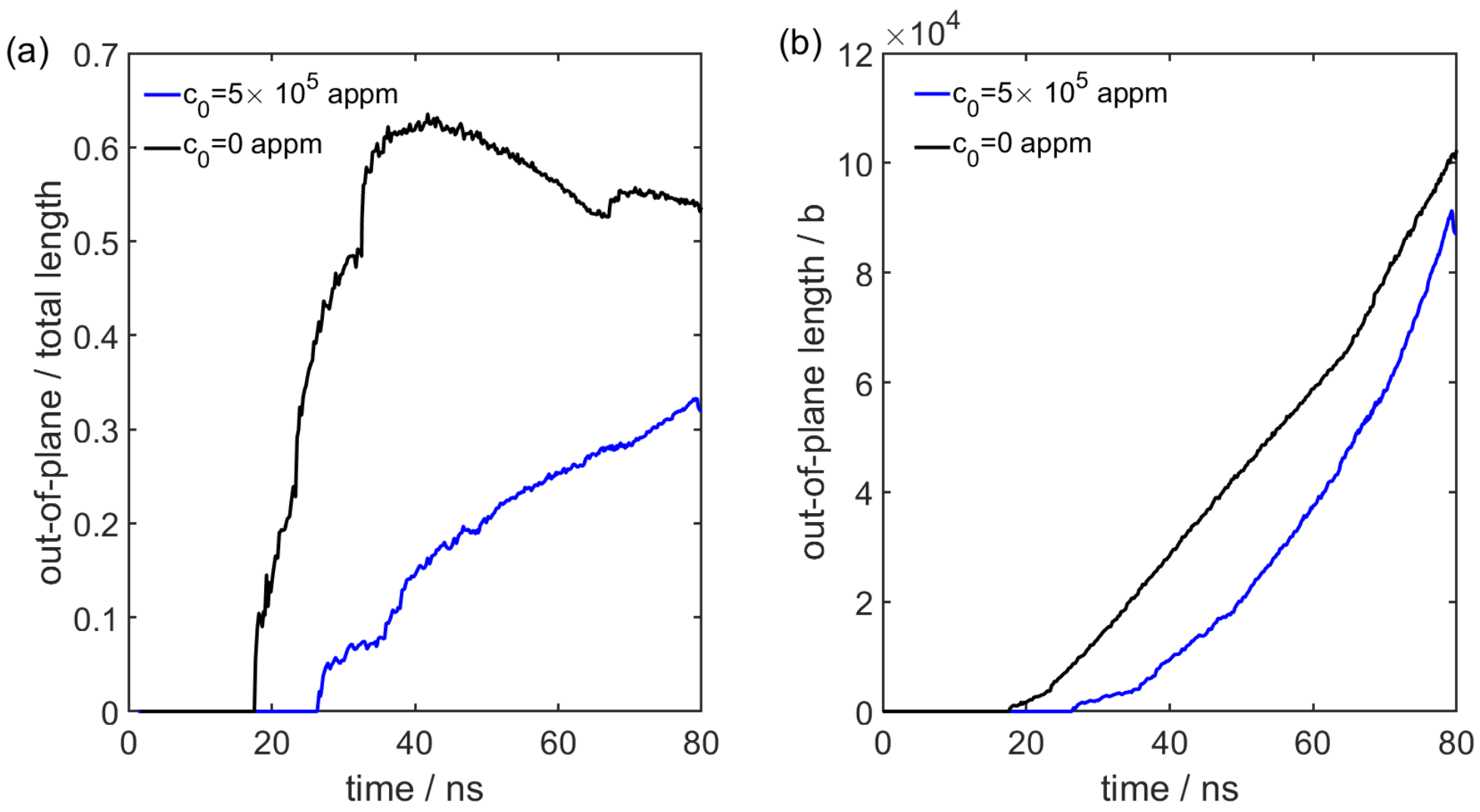}
\caption{Time histories of (a) the proportion of out-of-plane segments in the total dislocation structure and (b) the total length of out-of-plane segments. The hydrogen-free case is marked black, and the hydrogen elastic shielding case is marked blue. (For interpretation of the references to color in this figure, the reader is referred to the web version of this article.)}
\label{fig:planarityshieldingoutplane}
\end{figure}

During each iteration, all segments are checked, and a segment is considered out-of-plane  if both its nodes are, $\lvert\bm{X}_i\cdot \bm{n}\rvert>0$.  The total dislocation line length and the proportion which is out of plane are both reduced by the hydrogen elastic force. The onset of cross slip is also delayed as shown in figure \autoref{fig:planarityshieldingoutplane}

\begin{figure}[!h]
\centering\includegraphics[width=0.6\linewidth]{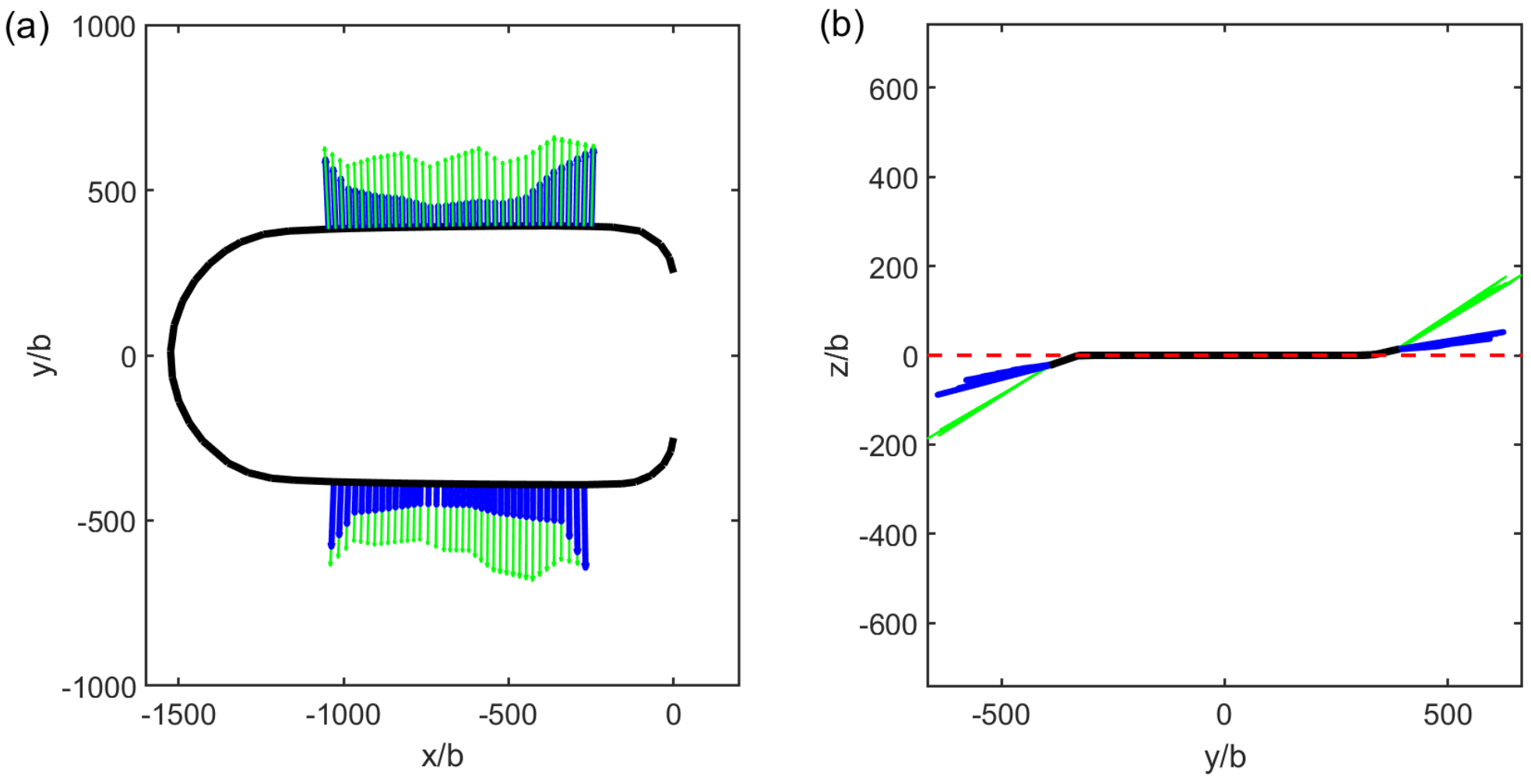}
\caption{ The nodal force partitioned into hydrogen, $\bar{\bm{F}}_k$ (blue arrows), and non-hydrogen parts $\tilde{\bm{F}}_k + \hat{\bm{F}}_k$ (green arrows) just after the first cross slip event. (a) shows the forces projected in the slip plane. (b) Shows the projection of forces normal to the slip plane, the red dashed line. Note $\bar{\bm{F}}_k$ has been scaled $10\times$ for clarity.}
\label{fig:planarityshieldingforce}
\end{figure}

The nodal force on screw like segments is shown in \autoref{fig:planarityshieldingforce}. 
For nodes on the slip plane, $\bar{\bm{F}}_k$ and $\tilde{\bm{F}}_k + \hat{\bm{F}}_k$ act in the same direction, driving the screw like segments apart but in opposition on the edge segments. $\lvert\tilde{\bm{F}}_k + \hat{\bm{F}}_k\rvert \gg \lvert \bar{\bm{F}}_k\rvert $ but still sufficient to decrease the anisotropy in the loop expansion, decreasing the screw length and hence the amount of cross-slip. This is consistent with the theory in \cite{Ferreira1999} that hydrogen tends to reduce the pure screw population.  $\bar{\bm{F}}_k$ has a small out of plane component but is predominantly in plane. Furthermore $\bar{\bm{F}}_k$ is entirely in plane prior to the initial out-of-plane motion which delays the first cross slip event. The hydrogen elastic force is only noticeable at extremely high bulk concentrations, and so can be neglected for bcc Fe.

\subsubsection{Hydrogen increased mobility}
\label{planaritymobility}

\begin{figure}[!h]
\centering\includegraphics[width=0.6\linewidth]{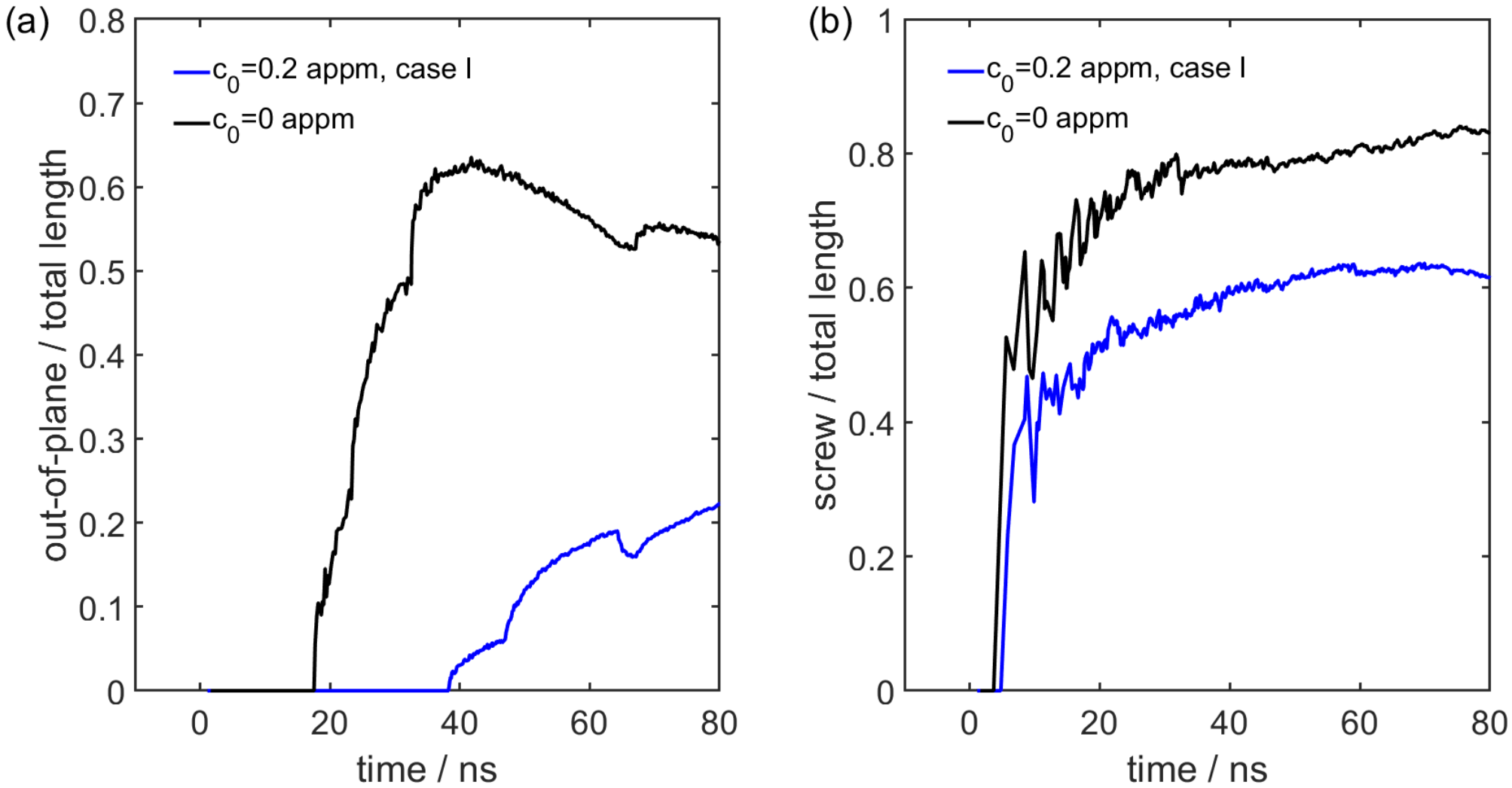}
\caption{Time histories of (a) the proportion of out-of-plane length and (b) the proportion of screw dominated length in the total dislocation population. The case without hydrogen is marked black, and that with hydrogen is marked blue.}
\label{fig:planarityHcaseI}
\end{figure}

The cross slip simulations were repeated with a hydrogen enhanced screw mobility law 
(case I) and a bulk concentration of $c_0=0.2$ appm. Time histories of the proportion of out-of-plane dislocation length are plotted in \autoref{fig:planarityHcaseI}(a).
The out-of-plane dislocation motion is delayed and suppressed in the presence of hydrogen. It is therfore hypothesised that  hydrogen enhanced slip planarity is due to the hydrogen induced shape change of dislocation loops. The evolution of the proportion of screw like segments is shown in \autoref{fig:planarityHcaseI}(b), which decreases in the presence of hydrogen. Indicating that hydrogen enhances slip planarity by reducing the amount of cross slip as the proportion of pure screw segments is reduced.

\begin{figure}[!h]
\centering\includegraphics[width=0.6\linewidth]{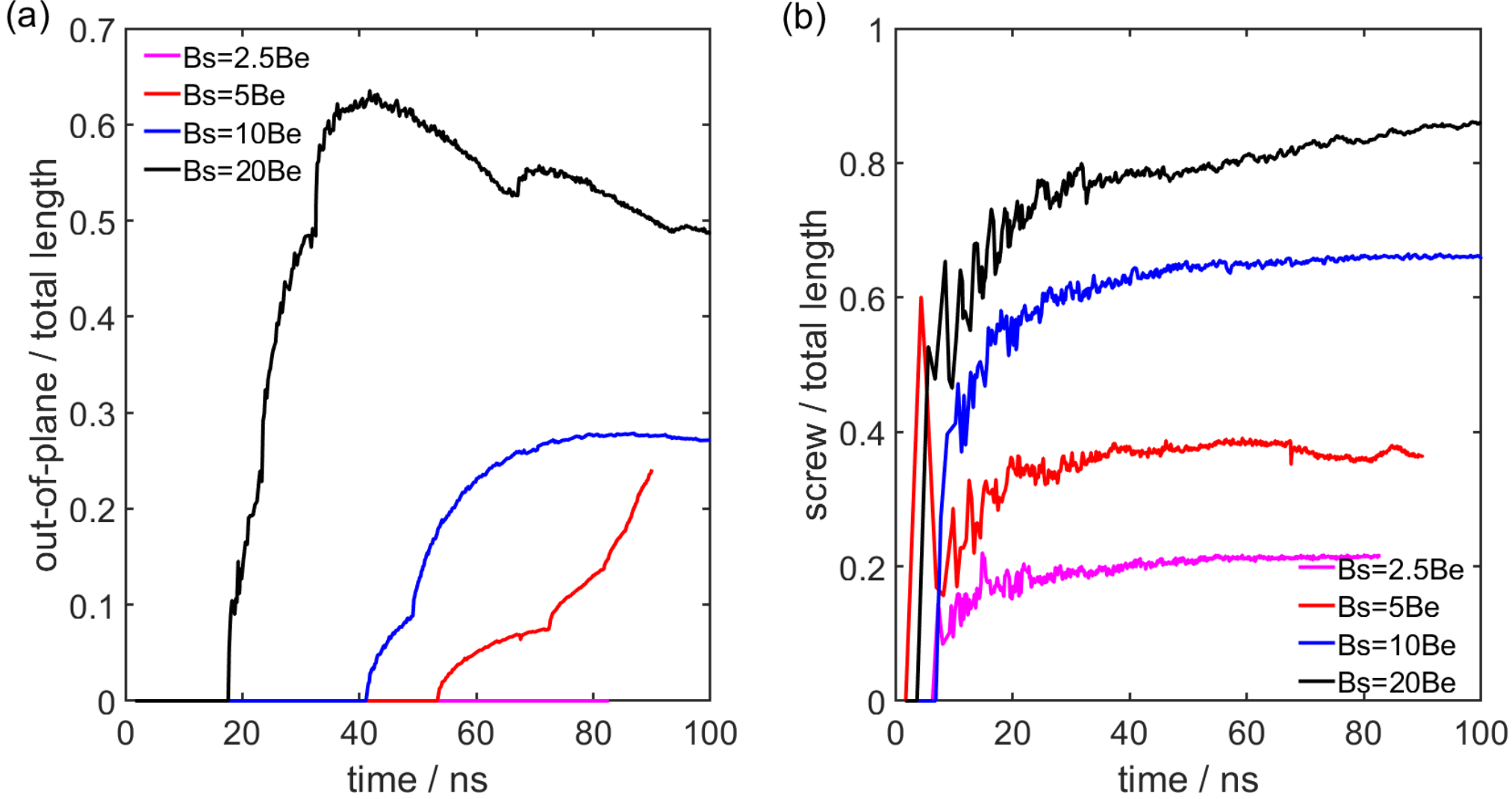}
\caption{Time histories of (a) the proportion of out-of-plane length and (b) the proportion of screw dominated length in the total dislocation population. Idealised simulations without hydrogen with the dislocation drag ratio $B_s/B_{eg}$ reduced from 20 down to 2.5.}
\label{fig:planarityscalemobility}
\end{figure}

To provide further support for this hypothesis, three simplified simulations with, $B_{s}\equiv 10B_{eg}$, $B_{s}\equiv 5B_{eg}$, and $B_{s}\equiv 2.5B_{eg}$ and no hydrogen, $c_0=0$, were performed. These ideal cases, shown in \autoref{fig:planarityscalemobility} qualitatively resemble the hydrogen enhanced screw mobility, \autoref{fig:planarityHcaseI}, and are free of the complexity associated with the hydrogen distribution. 

Hydrogen can therefore enhance slip planarity through hydrogen enhanced screw  mobility at a realistic bulk concentration $c_0=0.2$ appm for bcc Fe, in contrast to the extremely high value of $c_0=5\times10^5$ appm required for the hydrogen elastic force. Finally, it should be reemphasized that out-of-plane dislocation motions in this work are due to cross slip of pure screw segments, and climb is considered. The influence of hydrogen on climb will add further complexity.

\section{Hydrogen effects in a microcantilever}
\label{sec:finite}

\begin{figure}[!h]
\centering\includegraphics[width=0.6\linewidth]{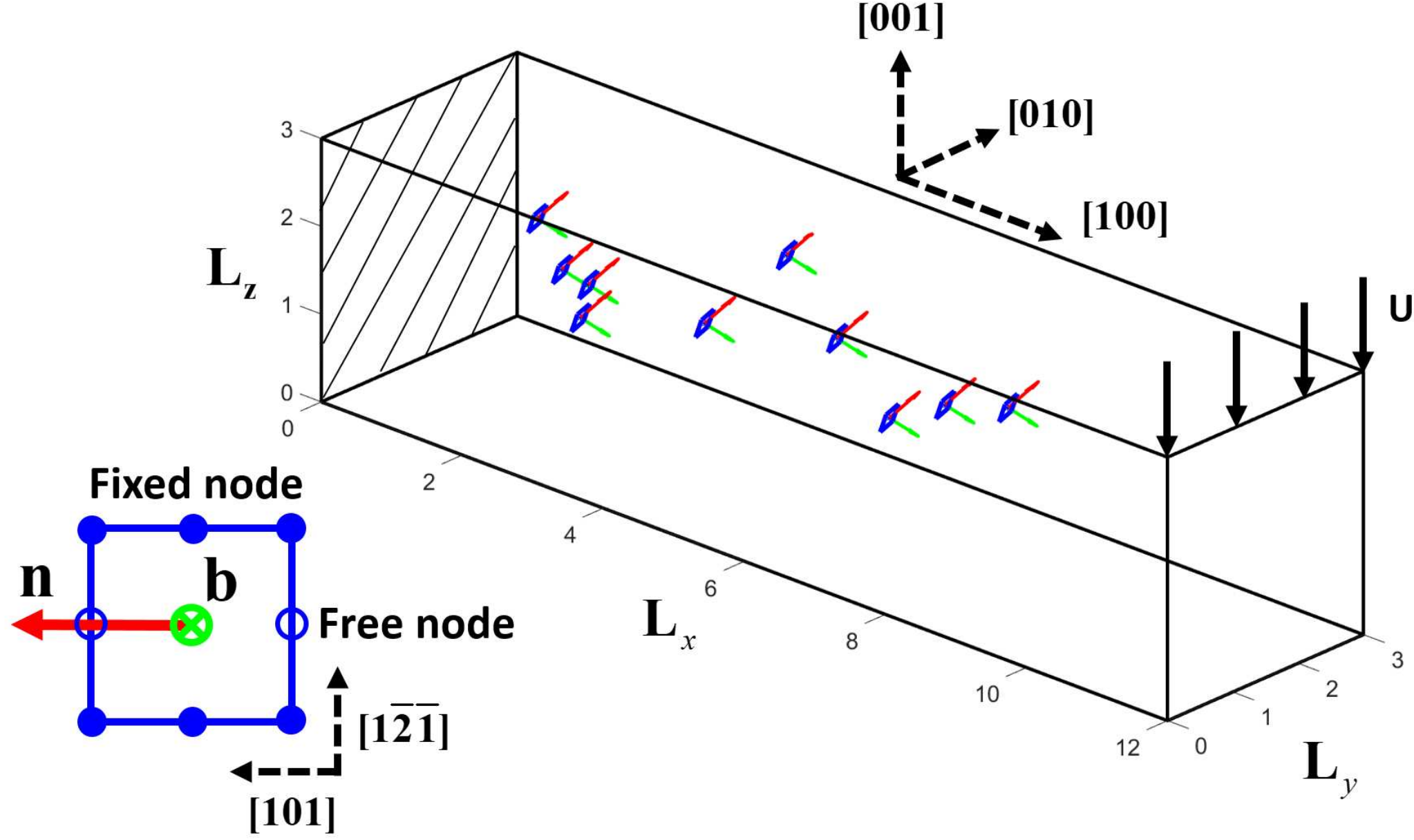}
\caption{Square prismatic loops with two free nodes are randomly generated on the $(101) [11\bar{1}]$ slip system as initial dislocation sources. The source (blue line) its Burgers vector (green arrow) and normal (red arrow) are shown.}
\label{fig:cantileverillustration}
\end{figure}

In situ hydrogen charged microcantilever bend tests on single crystal Fe-$3\%$ Si alloy and single crystal FeAl alloy were recently performed by \citet{Tarlan2017} and \citet{Yun2018}, respectively. Here we simulate a microcantilever with dimension  $L_x=12~\mu\textrm{m}$ and $L_y=L_z=3~\mu\textrm{m}$ to investigation hydrogen effects in a finite volume. The crystal orientation is aligned with the $\langle 100\rangle$ axis along the model axis as shown in \autoref{fig:cantileverillustration}; the same orientation used in the experiments by \citet{Yun2018}. However we do not attempt to reproduce the exact experimental data at the present stage, since the microcantilever geometry simulated here is simplified, with a square cross section and without a base or notch. For simplicity only the $(101) [11\bar{1}]$ slip system is included in the simulation. The sources were square prismatic loops with eight nodes, $6$ of which were fixed, as shown in \autoref{fig:cantileverillustration}. A total of 10 sources were used. The boundary conditions were: $\bm{u}(0,y,z)=[0,0,0]$,  $\bm{u}(L_x,y,L_z) = [0,0,U]$ with an applied displacement of $U=-0.169~\mu\textrm{m}$ and the remaining surfaces were traction free. Fully integrated linear brick finite elements, $0.24\ \mu m$ in size, were used to solve for the corrective elastic fields at every time increment.

\begin{figure}[!h]
\centering\includegraphics[width=0.6\linewidth]{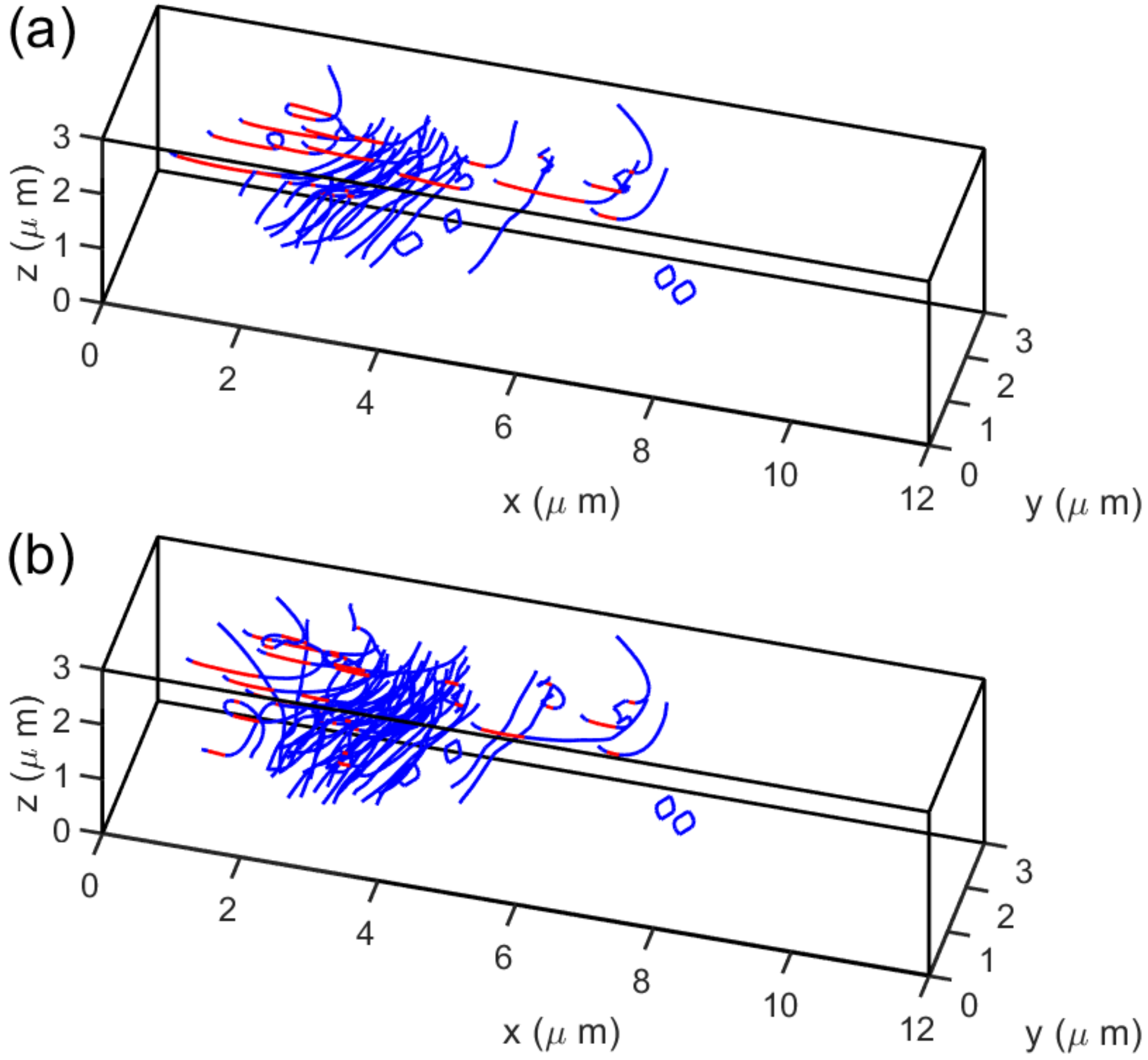}
\caption{Dislocations structures at $U=-0.169~\mu\textrm{m}$ for (a) the hydrogen free case and (b) the case with hydrogen increased screw mobility. The screw dominant segments are shown in red and all other segments in blue.}
\label{fig:structuresingle}
\end{figure}

\begin{figure}[!h]
\centering\includegraphics[width=0.6\linewidth]{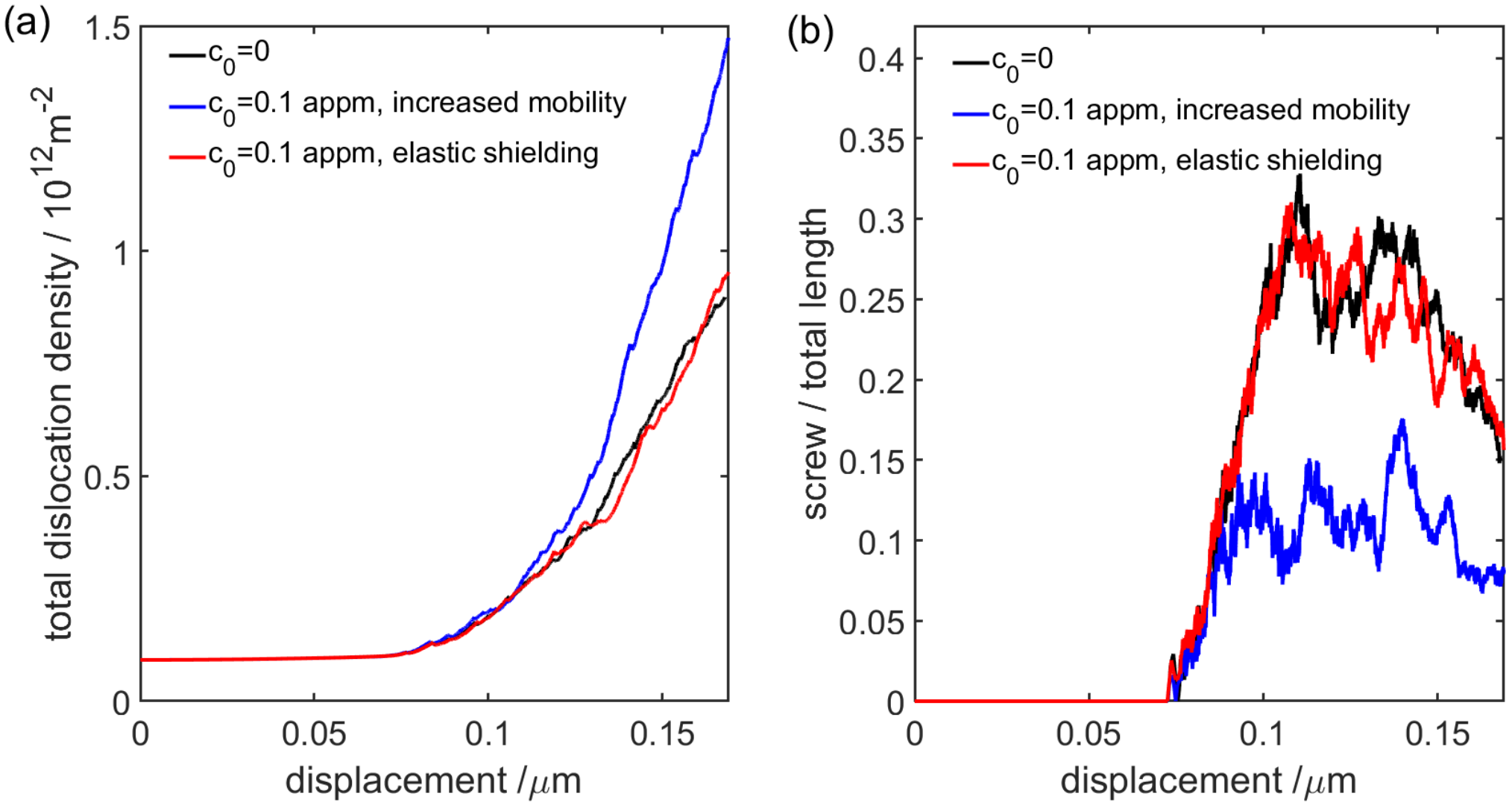}
\caption{(a) The dislocation density versus displacement curves for and (b) the proportion of screw dominant segments. The effect of the hydrogen elastic force is negligble.}
\label{fig:curvessingle}
\end{figure}

Hydrogen elastic shielding and hydrogen enhanced screw mobility are considered with a realistic initial bulk hydrogen concentration for bcc Fe, $c_0=0.1$ appm. The sources are subjected to an approximately uniaxial stress, $\sigma_{xx}$, which is tensile above and compressive below the neutral plane located at $z=1.5~\mu\textrm{m}$. The initial stress state is therefore similar to the slip planarity investigation in an infinite volume discussed in \autoref{Planarityresults}.
\begin{figure}[!h]
\centering\includegraphics[width=0.6\linewidth]{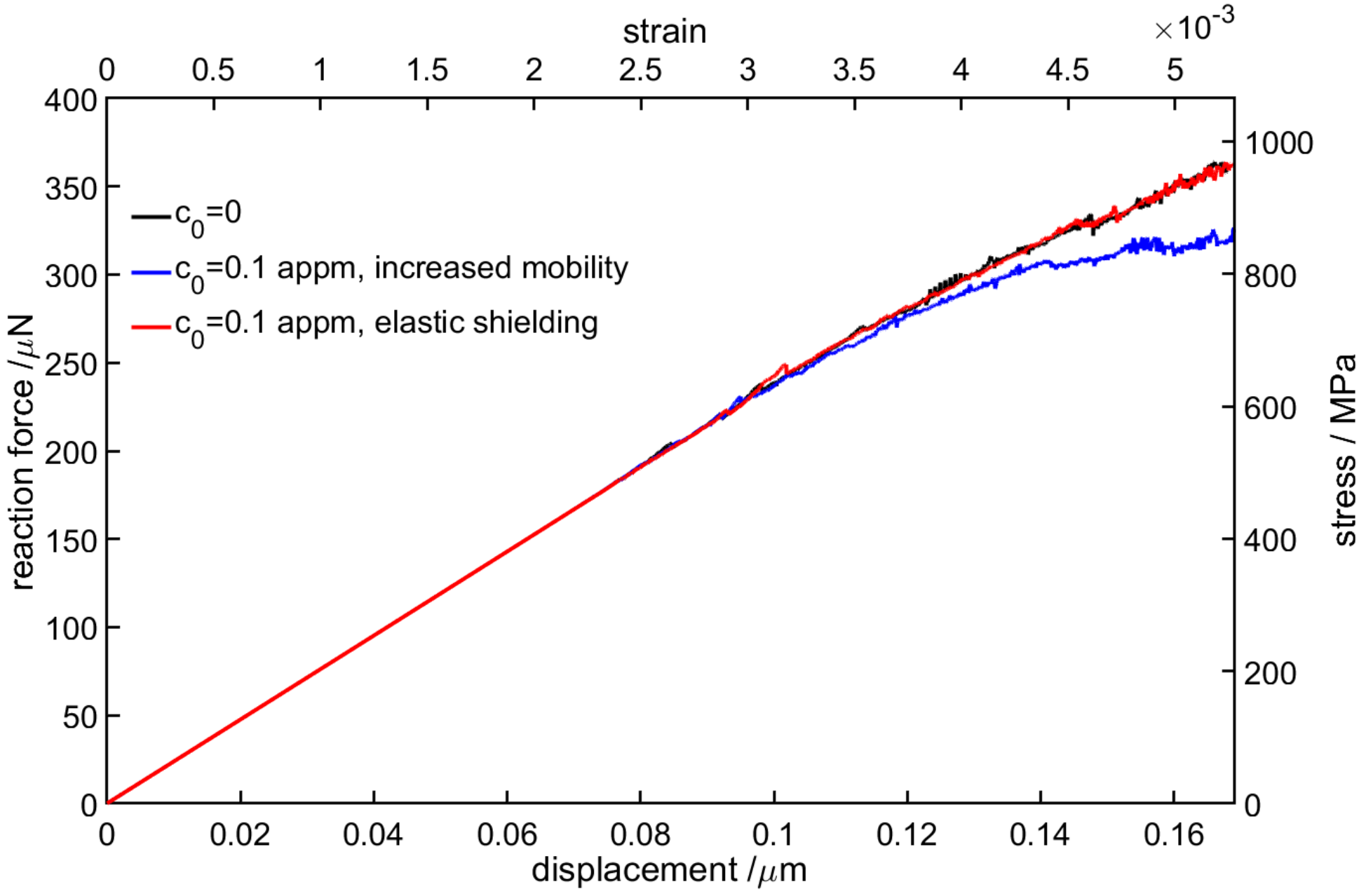}
\caption{Load displacement curve for the three cases considered.}
\label{fig:globalsingle}
\end{figure}
The load displacement curve is shown in \autoref{fig:globalsingle} which shows a clear softening effect through the hydrogen enhanced screw mobility law and no noticeable effect through hydrogen elastic shielding; consistent with the infinite volume simulations. The hydrogen softening can be rationalised by analyses of the dislocation structure, which is shown in \autoref{fig:structuresingle}.
In both cases, dislocation pile-ups form at the the neutral plane. However, the number of dislocations in a pile up is higher when hydrogen is included, and the proportion of screw like segments is reduced, as shown in \autoref{fig:curvessingle}. It is therefore concluded that hydrogen enhances dislocation activity in bcc Fe and consequently leads to global softening behavior, which is consistent with the HELP mechanism and agrees with the observation of hydrogen reducing the yield stress in microcantilever experiments \citep{Tarlan2017,Yun2018}. Our simulations indicate the effect is due to hydrogen increased screw mobility, and the contribution of hydrogen elastic shielding is negligible. Further, \autoref{fig:curvessingle}(b) suggests that hydrogen enhances slip planarity due to a reduction in the proportion of screw segments available to cross slip. The simulated dislocation density increases in the presence of hydrogen, this is in contrast to the expectation in some literature \citep{Nibur2006} that enhanced planarity would lead to hardening. This could be because only one slip system was simulated and so junction formation was not considered. Or it could be due to the existence of a neutral plane which arrests a large number of  dislocations. Multiple slip systems and an additional geometry, e.g. micropillar, with a uniform stress state will be simulated in future work to provide further insight. 

\begin{figure}[!h]
\centering\includegraphics[width=0.9\linewidth]{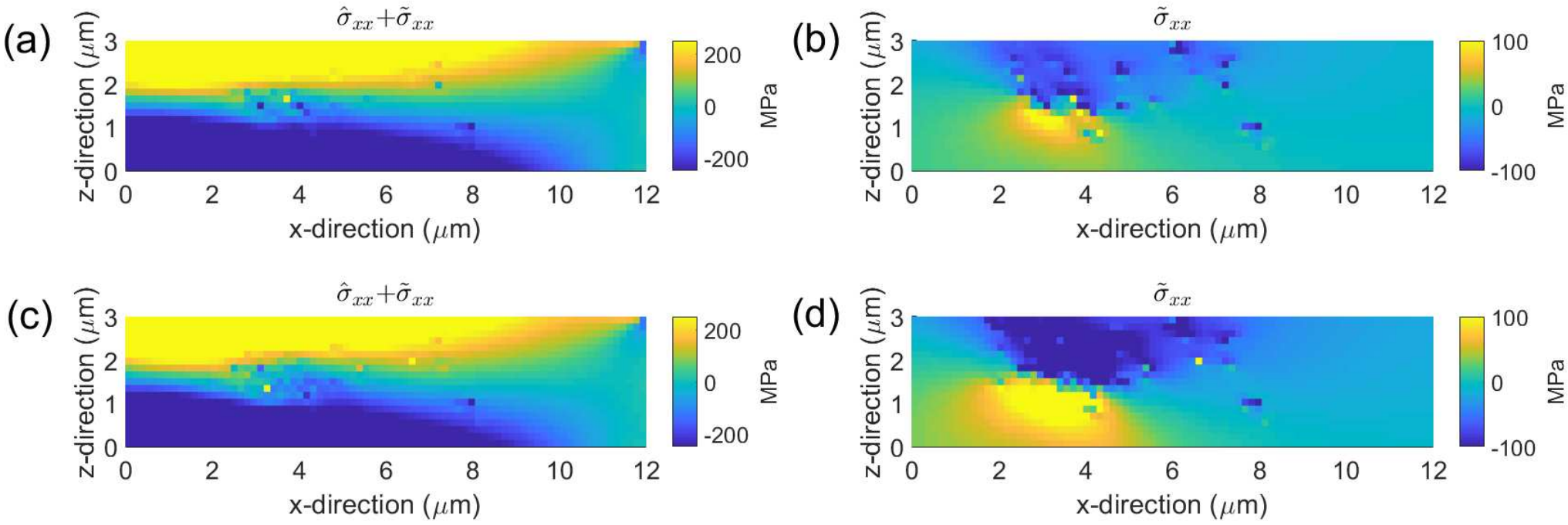}
\caption{(a) The total axial stress field $\hat{\sigma}_{xx}+\tilde{\sigma}_{xx}$ with $c_0=0$; (b) the axial stress field $\tilde{\sigma}_{xx}$ due to dislocations with $c_0=0$; (c) the total axial stress field $\hat{\sigma}_{xx}+\tilde{\sigma}_{xx}$ with $c_0=0.1$ appm; (d) the axial stress field $\tilde{\sigma}_{xx}$ due to dislocations with $c_0=0.1$ appm. Only hydrogen increased screw dislocation mobility is considered.}
\label{fig:stresssingle}
\end{figure}

The component of the dislocation stress, $\tilde{\sigma}_{xx}$, and the total stress, $\tilde{\sigma}_{xx}+\hat{\sigma}_{xx}$, along the beam axis, with and without hydrogen enhanced screw mobility are shown in \autoref{fig:stresssingle}. 
The stress hot spot is due to the pile up of edge dislocations at the neutral plane, and is more pronounced with hydrogen due to the higher dislocation density. As the dislocation field is compressive above the neutral plane and compressive below, the overall stress and reaction force is reduced more when hydrogen is present. 
\begin{figure}[!h]
\centering\includegraphics[width=0.6\linewidth]{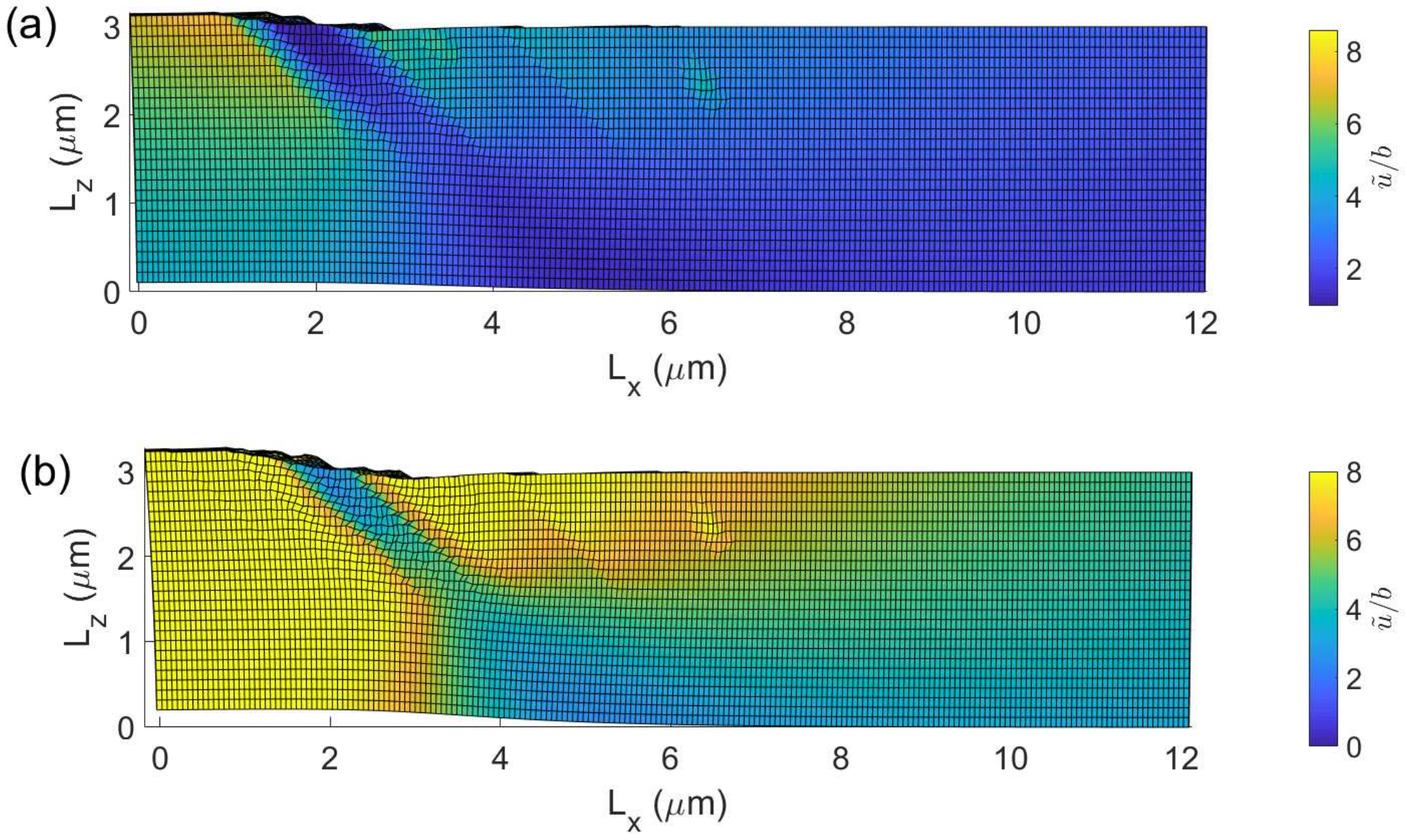}
\caption{The dislocation displacement field, $\tilde{\bm{u}}$, with (a) $c_0=0$ and (b) $c_0=0.1$ appm. The localised deformation is caused by dislocations that have exited the free surface; note that $\tilde{\bm{u}}$ has been scaled $100\times$ for clarity. The colour indicates the magnitude of the displacement $\tilde{{u}}=\lvert\tilde{\bm{u}}\rvert$}
\label{fig:deforsingle}
\end{figure}
The dislocation displacement field, $\tilde{\bm{u}}$, is shown in figure \autoref{fig:deforsingle}. When a dislocation segment leaves the material at a free surface it generates a plastic slip step. Hydrogen promotes plastic deformation, and the number of internal and exited segments increases, which increases the magnitude of the $\tilde{\bm{u}}$ field and the slip step height respectively.




\section{Conclusions}
\label{conclusion}
In this paper, the formulation proposed by \cite{Gu2018} to incorporate the hydrogen elastic force was implemented in DDD to simulate a Frank-Read source and within DDP to simulate a microcantilever. A hydrogen enhanced mobility law for screw dislocations was also implemented; based on hydrogen-kink pair interactions simulated by \cite{Katzarov2017} using kMC.

Hydrogen elastic shielding was found to have a tendency to promote dislocation generation, decrease dislocation spacing and enhance slip planarity. The elastic hydrogen force accelerates the bowing back and annihilation of screw dipoles in FR sources reduces the proportion of screw segments, which limits cross slip. These results are consistent with the HELP mechanism \citep{Birnbaum1994,Ferreira1999}. However our simulations demonstrated that this only occurs with an nonphysically high bulk hydrogen concentration, e.g. $c_0=1\times10^5$ appm in \cite{Gu2018} and $c_0=5\times10^5$ appm in this work, which is unrealistic for bcc metals. In contrast, a significant hydrogen influence on dislocation activity was observed by implementing a hydrogen concentration dependent mobility law for the screw component of dislocation segments. At realistic bulk hydrogen concentrations $c_0=0.1$ appm and $c_0=0.2$ appm hydrogen was found to promotes dislocation generation and enhances slip planarity. Hydrogen reduces the anisotropy in the mobility resulting in more circular loops \citep{Bhatia2014,Moriya1979}, and this also tends to reduce the screw population, making the bowing back of FR sources easier and reducing cross slip.

Both hydrogen elastic shielding and hydrogen increased screw mobility tend to enhance dislocation activity, and the mechanisms of the enhancement are actually quite similar. Therefore, hydrogen increased mobility should be viewed as a complementary explanation for the HELP mechanism, and it contributes a larger part than hydrogen (linear) elastic shielding, since it essentially arises from hydrogen-kink pair interactions close to the dislocation core, where very high stress and hydrogen concentrations exist but cannot be captured by linear elasticity theory.

Microcantilever simulations with a hydrogen dependent mobility law showed more pronounced dislocation pile-ups form at the neutral plane, and a reduction in the flow stress. Finally, it should be noted again that dislocation climb was excluded in all the simulations, therefore, hydrogen enhanced slip planarity is a consequence only of hydrogen affected cross slipping in the present work.

\section*{Acknowledgements}
This work was supported by the Engineering and Physical Sciences Research Council (Programme grant EP/L014742/1 and Fellowship grant EP/N007239/1). We also thank Prof. El-Awady for a useful discussion at the 3rd Sch{\"o}ental Symposium on Dislocation based Plasticity.




\section*{References}
\bibliographystyle{elsart-num-names}
\bibliography{HHDDDHELP.bib}

\begin{thebibliography}{10}
\expandafter\ifx\csname url\endcsname\relax
  \def\url#1{\texttt{#1}}\fi
\expandafter\ifx\csname urlprefix\endcsname\relax\def\urlprefix{URL }\fi

\bibitem[{Gu and El-Awady(2018)}]{Gu2018}
Y.~Gu, J.~A. El-Awady, Quantifying the effect of hydrogen on dislocation
  dynamics: A three-dimensional discrete dislocation dynamics framework,
  Journal of the Mechanics and Physics of Solids.

\bibitem[{Katzarov et~al.(2017)Katzarov, Pashov, and Paxton}]{Katzarov2017}
I.~H. Katzarov, D.~L. Pashov, A.~T. Paxton, Hydrogen embrittlement i. analysis
  of hydrogen-enhanced localized plasticity: Effect of hydrogen on the velocity
  of screw dislocations in $\ensuremath{\alpha}$-fe, Physical Review Materials
  1~(3) (2017) 033602, pRMATERIALS.

\bibitem[{Ayas et~al.(2014)Ayas, van Dommelen, and Deshpande}]{Ayas2014}
C.~Ayas, J.~A.~W. van Dommelen, V.~S. Deshpande, Climb-enabled discrete
  dislocation plasticity, Journal of the Mechanics and Physics of Solids 62
  (2014) 113--136.

\bibitem[{Robertson et~al.(2015)Robertson, Sofronis, Nagao, Martin, Wang,
  Gross, and Nygren}]{Robertson2015}
I.~Robertson, P.~Sofronis, A.~Nagao, M.~L. Martin, S.~Wang, D.~W. Gross, K.~E.
  Nygren, Hydrogen embrittlement understood, Metallurgical and Materials
  Transactions B 46~(3) (2015) 1085--1103.

\bibitem[{Oriani(1972)}]{Oriani1972}
R.~A. Oriani, A mechanistic theory of hydrogen embrittlement of steels,
  Berichte der Bunsengesellschaft für physikalische Chemie 76~(8) (1972)
  848--857.

\bibitem[{Lynch(2012)}]{Lynch2012}
S.~Lynch, Hydrogen embrittlement phenomena and mechanisms, Corrosion Reviews
  30~(3-4) (2012) 105.

\bibitem[{Birnbaum and Sofronis(1994)}]{Birnbaum1994}
H.~K. Birnbaum, P.~Sofronis, Hydrogen-enhanced localized plasticity—a
  mechanism for hydrogen-related fracture, Materials Science and Engineering: A
  176~(1) (1994) 191--202.

\bibitem[{Robertson(2001)}]{Robertson2001}
I.~M. Robertson, The effect of hydrogen on dislocation dynamics, Engineering
  Fracture Mechanics 68~(6) (2001) 671--692.

\bibitem[{Novak et~al.(2010)Novak, Yuan, Somerday, Sofronis, and
  Ritchie}]{Novak2010}
P.~Novak, R.~Yuan, B.~P. Somerday, P.~Sofronis, R.~O. Ritchie, A statistical,
  physical-based, micro-mechanical model of hydrogen-induced intergranular
  fracture in steel, Journal of the Mechanics and Physics of Solids 58~(2)
  (2010) 206--226.

\bibitem[{Barrera et~al.(2014)Barrera, Tarleton, and Cocks}]{Barrera2014}
O.~Barrera, E.~Tarleton, A.~Cocks, {A micromechanical image-based model for the
  featureless zone of a Fe–Ni dissimilar weld}, Philosophical Magazine
  94~(12) (2014) 1361--1377.

\bibitem[{Nagao et~al.(2018)Nagao, Dadfarnia, Somerday, Sofronis, and
  Ritchie}]{Nagao2018}
A.~Nagao, M.~Dadfarnia, B.~P. Somerday, P.~Sofronis, R.~O. Ritchie,
  Hydrogen-enhanced-plasticity mediated decohesion for hydrogen-induced
  intergranular and “quasi-cleavage” fracture of lath martensitic steels,
  Journal of the Mechanics and Physics of Solids 112 (2018) 403--430.

\bibitem[{Nagao et~al.(2012)Nagao, Smith, Dadfarnia, Sofronis, and
  Robertson}]{Nagao2012}
A.~Nagao, C.~D. Smith, M.~Dadfarnia, P.~Sofronis, I.~M. Robertson, The role of
  hydrogen in hydrogen embrittlement fracture of lath martensitic steel, Acta
  Materialia 60~(13–14) (2012) 5182--5189.

\bibitem[{Ferreira et~al.(1998)Ferreira, Robertson, and
  Birnbaum}]{Ferreira1998}
P.~J. Ferreira, I.~M. Robertson, H.~K. Birnbaum, Hydrogen effects on the
  interaction between dislocations, Acta Materialia 46~(5) (1998) 1749--1757.

\bibitem[{Shih et~al.(1988)Shih, Robertson, and Birnbaum}]{Shin1998}
D.~S. Shih, I.~M. Robertson, H.~K. Birnbaum, Hydrogen embrittlement of $\alpha$
  titanium: In situ tem studies, Acta Metallurgica 36~(1) (1988) 111--124.

\bibitem[{Tabata and Birnbaum(1983)}]{Tabata1983}
T.~Tabata, H.~Birnbaum, Direct observations of the effect of hydrogen on the
  behavior of dislocations in iron, Scripta Metallurgica 17~(7) (1983)
  947--950.

\bibitem[{Robertson and Birnbaum(1986)}]{Robertson1986}
I.~M. Robertson, H.~K. Birnbaum, An hvem study of hydrogen effects on the
  deformation and fracture of nickel, Acta Metallurgica 34~(3) (1986) 353--366.

\bibitem[{Bond et~al.(1988)Bond, Robertson, and Birnbaum}]{Bond1988}
G.~M. Bond, I.~M. Robertson, H.~K. Birnbaum, Effects of hydrogen on deformation
  and fracture processes in high-purity aluminium, Acta Metallurgica 36~(8)
  (1988) 2193--2197.

\bibitem[{Rozenak et~al.(1990)Rozenak, Robertson, and Birnbaum}]{Rozenak1990}
P.~Rozenak, I.~M. Robertson, H.~K. Birnbaum, Hvem studies of the effects of
  hydrogen on the deformation and fracture of aisi type 316 austenitic
  stainless steel, Acta Metallurgica et Materialia 38~(11) (1990) 2031--2040.

\bibitem[{Bond et~al.(1989)Bond, Robertson, and Birnbaum}]{Bond1989}
G.~M. Bond, I.~M. Robertson, H.~K. Birnbaum, On the mechanisms of hydrogen
  embrittlement of ni3al alloys, Acta Metallurgica 37~(5) (1989) 1407--1413.

\bibitem[{Ferreira et~al.(1999)Ferreira, Robertson, and
  Birnbaum}]{Ferreira1999}
P.~J. Ferreira, I.~M. Robertson, H.~K. Birnbaum, Hydrogen effects on the
  character of dislocations in high-purity aluminum, Acta Materialia 47~(10)
  (1999) 2991--2998.

\bibitem[{Nibur et~al.(2006)Nibur, Bahr, and Somerday}]{Nibur2006}
K.~A. Nibur, D.~F. Bahr, B.~P. Somerday, Hydrogen effects on dislocation
  activity in austenitic stainless steel, Acta Materialia 54~(10) (2006)
  2677--2684.

\bibitem[{Barnoush and Vehoff(2010)}]{Afrooz2010}
A.~Barnoush, H.~Vehoff, Recent developments in the study of hydrogen
  embrittlement: Hydrogen effect on dislocation nucleation, Acta Materialia
  58~(16) (2010) 5274--5285.

\bibitem[{Hajilou et~al.(2017)Hajilou, Deng, Rogne, Kheradmand, and
  Barnoush}]{Tarlan2017}
T.~Hajilou, Y.~Deng, B.~R. Rogne, N.~Kheradmand, A.~Barnoush, In situ
  electrochemical microcantilever bending test: A new insight into hydrogen
  enhanced cracking, Scripta Materialia 132 (2017) 17--21.

\bibitem[{Deng et~al.(2017)Deng, Hajilou, Wan, Kheradmand, and
  Barnoush}]{Yun2017}
Y.~Deng, T.~Hajilou, D.~Wan, N.~Kheradmand, A.~Barnoush, In-situ
  micro-cantilever bending test in environmental scanning electron microscope:
  Real time observation of hydrogen enhanced cracking, Scripta Materialia 127
  (2017) 19--23.

\bibitem[{Deng and Barnoush(2018)}]{Yun2018}
Y.~Deng, A.~Barnoush, Hydrogen embrittlement revealed via novel in situ
  fracture experiments using notched micro-cantilever specimens, Acta
  Materialia 142 (2018) 236--247.

\bibitem[{Ulmer and Altstetter(1991)}]{Ulmer1991}
D.~G. Ulmer, C.~J. Altstetter, Hydrogen-induced strain localization and failure
  of austenitic stainless steels at high hydrogen concentrations, Acta
  Metallurgica et Materialia 39~(6) (1991) 1237--1248.

\bibitem[{Matsuo et~al.(2014)Matsuo, Yamabe, and Matsuoka}]{Takashi2014}
T.~Matsuo, J.~Yamabe, S.~Matsuoka, Effects of hydrogen on tensile properties
  and fracture surface morphologies of type 316l stainless steel, International
  Journal of Hydrogen Energy 39~(7) (2014) 3542--3551.

\bibitem[{Jagodzinski et~al.(2000)Jagodzinski, Hänninen, Tarasenko, and
  Smuk}]{Jagodzinski2000}
Y.~Jagodzinski, H.~Hänninen, O.~Tarasenko, S.~Smuk, Interaction of hydrogen
  with dislocation pile-ups and hydrogen induced softening of pure iron,
  Scripta Materialia 43~(3) (2000) 245--251.

\bibitem[{Wen et~al.(2003)Wen, Fukuyama, and Yokogawa}]{Wen2003}
M.~Wen, S.~Fukuyama, K.~Yokogawa, Atomistic simulations of effect of hydrogen
  on kink-pair energetics of screw dislocations in bcc iron, Acta Materialia
  51~(6) (2003) 1767--1773.

\bibitem[{Teus et~al.(2007)Teus, Shivanyuk, Shanina, and Gavriljuk}]{Teus2007}
S.~M. Teus, V.~N. Shivanyuk, B.~D. Shanina, V.~G. Gavriljuk, Effect of hydrogen
  on electronic structure of fcc iron in relation to hydrogen embrittlement of
  austenitic steels, physica status solidi (a) 204~(12) (2007) 4249--4258.

\bibitem[{Taketomi et~al.(2008)Taketomi, Matsumoto, and
  Miyazaki}]{Taketomi2008}
S.~Taketomi, R.~Matsumoto, N.~Miyazaki, Atomistic simulation of the effects of
  hydrogen on the mobility of edge dislocation in alpha iron, Journal of
  Materials Science 43~(3) (2008) 1166--1169.

\bibitem[{Gavriljuk et~al.(2010)Gavriljuk, Shanina, Shyvanyuk, and
  Teus}]{Gavriljuk2010}
V.~G. Gavriljuk, B.~D. Shanina, V.~N. Shyvanyuk, S.~M. Teus, Electronic effect
  on hydrogen brittleness of austenitic steels, Journal of Applied Physics
  108~(8) (2010) 083723.

\bibitem[{Song and Curtin(2011)}]{Song2011}
J.~Song, W.~A. Curtin, A nanoscale mechanism of hydrogen embrittlement in
  metals, Acta Materialia 59~(4) (2011) 1557--1569.

\bibitem[{Song and Curtin(2013)}]{Song2013}
J.~Song, W.~Curtin, Atomic mechanism and prediction of hydrogen embrittlement
  in iron, Nature materials 12~(2) (2013) 145.

\bibitem[{Song and Curtin(2014)}]{Song2014}
J.~Song, W.~A. Curtin, Mechanisms of hydrogen-enhanced localized plasticity: An
  atomistic study using $\alpha$-fe as a model system, Acta Materialia 68
  (2014) 61--69.

\bibitem[{Bhatia et~al.(2014)Bhatia, Groh, and Solanki}]{Bhatia2014}
M.~A. Bhatia, S.~Groh, K.~N. Solanki, Atomic-scale investigation of point
  defects and hydrogen-solute atmospheres on the edge dislocation mobility in
  alpha iron, Journal of Applied Physics 116~(6) (2014) 064302.

\bibitem[{Anderson et~al.(2017)Anderson, Hirth, and Lothe}]{anderson2017}
P.~Anderson, J.~Hirth, J.~Lothe, Theory of Dislocations, Cambridge University
  Press, 2017.

\bibitem[{Itakura et~al.(2013)Itakura, Kaburaki, Yamaguchi, and
  Okita}]{Itakura2013}
M.~Itakura, H.~Kaburaki, M.~Yamaguchi, T.~Okita, The effect of hydrogen atoms
  on the screw dislocation mobility in bcc iron: A first-principles study, Acta
  Materialia 61~(18) (2013) 6857--6867.

\bibitem[{Zhao and Lu(2011)}]{Zhao2011}
Y.~Zhao, G.~Lu, Qm/mm study of dislocation-hydrogen/helium interactions in
  $\alpha$-fe, Modelling and Simulation in Materials Science and Engineering
  19~(6) (2011) 065004.

\bibitem[{Moriya et~al.(1979)Moriya, Matsui, and Kimura}]{Moriya1979}
S.~Moriya, H.~Matsui, H.~Kimura, The effect of hydrogen on the mechanical
  properties of high purity iron ii. effect of quenched-in hydrogen below room
  temperature, Materials Science and Engineering 40~(2) (1979) 217--225.

\bibitem[{Sofronis et~al.(2001)Sofronis, Liang, and Aravas}]{Sofronis2001}
P.~Sofronis, Y.~Liang, N.~Aravas, Hydrogen induced shear localization of the
  plastic flow in metals and alloys, European Journal of Mechanics - A/Solids
  20~(6) (2001) 857--872.

\bibitem[{Liang et~al.(2003)Liang, Sofronis, and Aravas}]{Liang2003}
Y.~Liang, P.~Sofronis, N.~Aravas, On the effect of hydrogen on plastic
  instabilities in metals, Acta Materialia 51~(9) (2003) 2717--2730.

\bibitem[{Ahn et~al.(2007)Ahn, Sofronis, and Dodds~Jr}]{Ahn2007}
D.~C. Ahn, P.~Sofronis, R.~H. Dodds~Jr, On hydrogen-induced plastic flow
  localization during void growth and coalescence, International Journal of
  Hydrogen Energy 32~(16) (2007) 3734--3742.

\bibitem[{Barrera et~al.(2016)Barrera, Tarleton, Tang, and Cocks}]{Barrera2016}
O.~Barrera, E.~Tarleton, H.~W. Tang, A.~C.~F. Cocks, Modelling the coupling
  between hydrogen diffusion and the mechanical behaviour of metals,
  Computational Materials Science 122 (2016) 219--228.

\bibitem[{Yu et~al.(2018)Yu, Olsen, He, and Zhang}]{Yu2018}
H.~Yu, J.~S. Olsen, J.~He, Z.~Zhang, Hydrogen-microvoid interactions at
  continuum scale, International Journal of Hydrogen Energy.

\bibitem[{Capolungo(2011)}]{Capolungo2011}
L.~Capolungo, Dislocation junction formation and strength in magnesium, Acta
  Materialia 59~(8) (2011) 2909--2917.

\bibitem[{Wu et~al.(2013)Wu, Chung, Aubry, Munday, and Arsenlis}]{Wu2013}
C.~C. Wu, P.~W. Chung, S.~Aubry, L.~B. Munday, A.~Arsenlis, The strength of
  binary junctions in hexagonal close-packed crystals, Acta Materialia 61~(9)
  (2013) 3422--3431.

\bibitem[{Arsenlis et~al.(2007)Arsenlis, Cai, Tang, Rhee, Oppelstrup, Hommes,
  Pierce, and Bulatov}]{Arsenlis2007}
A.~Arsenlis, W.~Cai, M.~Tang, M.~Rhee, T.~Oppelstrup, G.~Hommes, T.~G. Pierce,
  V.~V. Bulatov, Enabling strain hardening simulations with dislocation
  dynamics, Modelling and Simulation in Materials Science and Engineering
  15~(6) (2007) 553.

\bibitem[{Fitzgerald et~al.(2012)Fitzgerald, Aubry, Dudarev, and
  Cai}]{Fitzgerald2012}
S.~P. Fitzgerald, S.~Aubry, S.~L. Dudarev, W.~Cai, Dislocation dynamics
  simulation of frank-read sources in anisotropic $\alpha$-fe, Modelling and
  Simulation in Materials Science and Engineering 20~(4) (2012) 045022.

\bibitem[{Aubry et~al.(2011)Aubry, Fitzgerald, Dudarev, and Cai}]{Aubry2011}
S.~Aubry, S.~P. Fitzgerald, S.~L. Dudarev, W.~Cai, Equilibrium shape of
  dislocation shear loops in anisotropic $\alpha$-fe, Modelling and Simulation
  in Materials Science and Engineering 19~(6) (2011) 065006.

\bibitem[{Motz et~al.(2008)Motz, Weygand, Senger, and Gumbsch}]{Motz2008}
C.~Motz, D.~Weygand, J.~Senger, P.~Gumbsch, Micro-bending tests: A comparison
  between three-dimensional discrete dislocation dynamics simulations and
  experiments, Acta Materialia 56~(9) (2008) 1942--1955.

\bibitem[{Motz et~al.(2009)Motz, Weygand, Senger, and Gumbsch}]{Motz2009}
C.~Motz, D.~Weygand, J.~Senger, P.~Gumbsch, Initial dislocation structures in
  3-d discrete dislocation dynamics and their influence on microscale
  plasticity, Acta Materialia 57~(6) (2009) 1744--1754.

\bibitem[{Ryu et~al.(2013)Ryu, Nix, and Cai}]{Ryu2013}
I.~Ryu, W.~D. Nix, W.~Cai, Plasticity of bcc micropillars controlled by
  competition between dislocation multiplication and depletion, Acta Materialia
  61~(9) (2013) 3233--3241.

\bibitem[{Po et~al.(2016)Po, Cui, Rivera, Cereceda, Swinburne, Marian, and
  Ghoniem}]{Po2016}
G.~Po, Y.~Cui, D.~Rivera, D.~Cereceda, T.~D. Swinburne, J.~Marian, N.~Ghoniem,
  A phenomenological dislocation mobility law for bcc metals, Acta Materialia
  119 (2016) 123--135.

\bibitem[{Cui et~al.(2016)Cui, Po, and Ghoniem}]{Cui2016}
Y.~Cui, G.~Po, N.~Ghoniem, Temperature insensitivity of the flow stress in
  body-centered cubic micropillar crystals, Acta Materialia 108 (2016)
  128--137.

\bibitem[{Cui et~al.(2017)Cui, Po, and Ghoniem}]{Cui2017}
Y.~Cui, G.~Po, N.~Ghoniem, Does irradiation enhance or inhibit strain bursts at
  the submicron scale?, Acta Materialia 132 (2017) 285--297.

\bibitem[{Cai et~al.(2014)Cai, Sills, Barnett, and Nix}]{Cai2014}
W.~Cai, R.~B. Sills, D.~M. Barnett, W.~D. Nix, Modeling a distribution of point
  defects as misfitting inclusions in stressed solids, Journal of the Mechanics
  and Physics of Solids 66 (2014) 154--171.

\bibitem[{Giessen and Needleman(1999)}]{Giessen1999}
E.~V.~D. Giessen, a.~Needleman, {Discrete dislocation plasticity: a simple
  planar model}, Modelling and Simulation in Materials Science and Engineering
  3 (1999) 689--735.

\bibitem[{El-Awady et~al.(2008)El-Awady, Bulent~Biner, and
  Ghoniem}]{Jaafar2008}
J.~A. El-Awady, S.~Bulent~Biner, N.~M. Ghoniem, A self-consistent boundary
  element, parametric dislocation dynamics formulation of plastic flow in
  finite volumes, Journal of the Mechanics and Physics of Solids 56~(5) (2008)
  2019--2035.

\bibitem[{Bulatov and Cai(2006)}]{Cai2006book}
V.~Bulatov, W.~Cai, Computer simulations of dislocations, Vol.~3, Oxford
  University Press on Demand, 2006.

\bibitem[{Cai et~al.(2006)Cai, Arsenlis, Weinberger, and Bulatov}]{Cai2006}
W.~Cai, A.~Arsenlis, C.~R. Weinberger, V.~V. Bulatov, A non-singular continuum
  theory of dislocations, Journal of the Mechanics and Physics of Solids 54~(3)
  (2006) 561--587.

\bibitem[{Cai and Bulatov(2004)}]{Cai2004}
W.~Cai, V.~V. Bulatov, Mobility laws in dislocation dynamics simulations,
  Materials Science and Engineering: A 387-389 (2004) 277--281.

\bibitem[{Giessen and Needleman(1995)}]{Needleman1995}
E.~V.~d. Giessen, A.~Needleman, Discrete dislocation plasticity: a simple
  planar model, Modelling and Simulation in Materials Science and Engineering
  3~(5) (1995) 689.

\bibitem[{Weygand et~al.(2002)Weygand, Friedman, Giessen, and
  Needleman}]{Weygand2002}
D.~Weygand, L.~H. Friedman, E.~V.~d. Giessen, A.~Needleman, Aspects of
  boundary-value problem solutions with three-dimensional dislocation dynamics,
  Modelling and Simulation in Materials Science and Engineering 10~(4) (2002)
  437.

\bibitem[{Eshelby(1955)}]{Eshelby1957}
J.~D. Eshelby, The elastic interaction of point defects, Acta Metallurgica
  3~(5) (1955) 487--490.

\bibitem[{Xie et~al.(2016)Xie, Li, Li, Wang, Gumbsch, Sun, Ma, Li, and
  Shan}]{Xie2016}
D.~Xie, S.~Li, M.~Li, Z.~Wang, P.~Gumbsch, J.~Sun, E.~Ma, J.~Li, Z.~Shan,
  Hydrogenated vacancies lock dislocations in aluminium, Nature communications
  7 (2016) 13341.

\bibitem[{Xie(25 May 2018)}]{Xie2018}
D.~Xie, Personal communication, 25 May 2018.

\bibitem[{Tang and Marian(2014)}]{Tang2014}
M.~Tang, J.~Marian, Temperature and high strain rate dependence of tensile
  deformation behavior in single-crystal iron from dislocation dynamics
  simulations, Acta Materialia 70 (2014) 123--129.

\bibitem[{Wang and Beyerlein(2011)}]{Wang2011}
Z.~Q. Wang, I.~J. Beyerlein, An atomistically-informed dislocation dynamics
  model for the plastic anisotropy and tension–compression asymmetry of bcc
  metals, International Journal of Plasticity 27~(10) (2011) 1471--1484.

\end{thebibliography}







\end{document}